\documentclass[%
 reprint,
 amsmath,
 amssymb,
 showkeys,
 pra,
 floatfix,
]{revtex4-2}

\usepackage{xr}

\usepackage[linesnumbered,ruled,vlined]{algorithm2e}
\let\oldnl\nl
\newcommand{\nonl}{\renewcommand{\nl}{\let\nl\oldnl}}
\usepackage{braket}
\usepackage{float}

\usepackage{graphicx}
\usepackage{dcolumn}
\usepackage{enumitem}
\usepackage{lipsum}  
\usepackage{bm}
\usepackage{hyperref}
\hypersetup{
    colorlinks=true,
    linkcolor=blue,
    citecolor=magenta
}
\usepackage{subcaption}
\begin{document}

\preprint{APS/123-QED}

\title{qLEET: Visualizing Loss Landscapes, Expressibility, Entangling power and Training Trajectories for Parameterized Quantum Circuits}

\author{Utkarsh Azad}
\email{utkarsh.azad@research.iiit.ac.in}
\thanks{Corresponding Author}
\author{Animesh Sinha}
\email{animesh.sinha@research.iiit.ac.in}

\affiliation{%
    Center for Computational Natural Sciences and Bioinformatics, International Institute of Information Technology, Hyderabad.\\
    Center for Quantum Science and Technology,\\ International Institute of Information Technology, Hyderabad.
}%

\date{\today}

\begin{abstract}

    We present qLEET, an open-source Python package for studying parameterized quantum circuits (PQCs), which are widely used in various variational quantum algorithms (VQAs) and quantum machine learning (QML) algorithms. qLEET enables computation of properties such as expressibility and entangling power of a PQC by studying its entanglement spectrum and the distribution of parameterized states produced by it. Furthermore, it allows users to visualize the training trajectories of PQCs along with high-dimensional loss landscapes generated by them for different objective functions. It supports quantum circuits and noise models built using popular quantum computing libraries such as Qiskit, Cirq, and Pyquil. In our work, we demonstrate how qLEET provides opportunities to design and improve hybrid quantum-classical algorithms by utilizing intuitive insights from the ansatz capability and structure of the loss landscape.

\end{abstract}

\maketitle


\section{\label{sec:intro}Introduction}

Recent advances in the field of quantum technologies have led to the development of near-term quantum hardware, more popularly referred to as noisy intermediate-scale quantum (NISQ) devices \cite{nisq_preskill, RevModPhys.94.015004}. Unfortunately, due to restrictive qubit connectivity, imperfect qubit control, and minimal error correction, their computation capabilities are limited to executing only low depth algorithms \cite{9251243}. For this reason, these devices are supposedly used as accelerators for their classical counterparts instead of stand-alone devices themselves. This has led to the development of hybrid quantum-classical (HQC) algorithms, which use both quantum and classical hardware either iteratively or sequentially. The problems are decomposed into classically tractable and intractable parts in such a setup, where the latter is solved using the quantum processor \cite{Endo2021-zy}. 

Parameterized quantum circuits (PQCs) are one of the fundamental components of these algorithms \cite{Benedetti2019-gz}. They are responsible for evolving the qubits system to a state which is dependent on the series of parameters ($\vec{\theta}$) provided by a classical processor and the objective function from some initial state $\ket{\psi_0}$. The initial state of the qubit system here could either be ground state $\ket{0\ldots0}$, or some other particular state such as Hartree-Fock state $\ket{\psi}_{HF}$ as in the case of electronic structure problems. The PQC ($U(\vec{\theta})$) is also popularly referred to as \textit{ansatz} \cite{Benedetti2019-gz}. Their structure dramatically affects the performance of HQCs as they influence both the (i) convergence speed, i.e., the number of quantum-classical feedback iterations, and (ii) closeness of the final state ($\ket{\psi({\vec{\theta}})}$) to a state that optimally solves the problem ($\ket{\psi({\vec{\theta}^*})}$), i.e., the overlap or the fidelity ($\mathcal{F} = |\langle\psi({\vec{\theta})} | \psi({\vec{\theta}^*})\rangle|^{2}$) \cite{9781107002173} between the final state and the target state. 

Therefore, it becomes imperative to design optimal PQCs for a given problem. However, this is not straightforward because their design depends not only on the problem instances themselves but also on the quantum hardware that executes them. After all, some essential properties like depth of circuit post compilation depend on the hardware's topology and the supported native gates. Overall, there exist three main classes of ansätze: (i) problem-inspired ansatz, where the evolutions of generators derived from properties of the given system are used to construct the PQCs \cite{Romero2018-uc}, (ii) hardware-efficient ansatz, where a minimal set of quantum gates native to a given device are used to construct the PQCs \cite{Kandala2017-wn}, and (iii) adaptive ansatz, which is midway between the former two ansätze \cite{PRXQuantum.2.020310}. Using these three classes, one can develop numerous ansatz designs for any given problem. However, to finally choose one, we need to have insights from the problems and a concrete strategy to compare their performances.

\begin{figure*}[!th]
    \centering
    \includegraphics[width=0.65\linewidth]{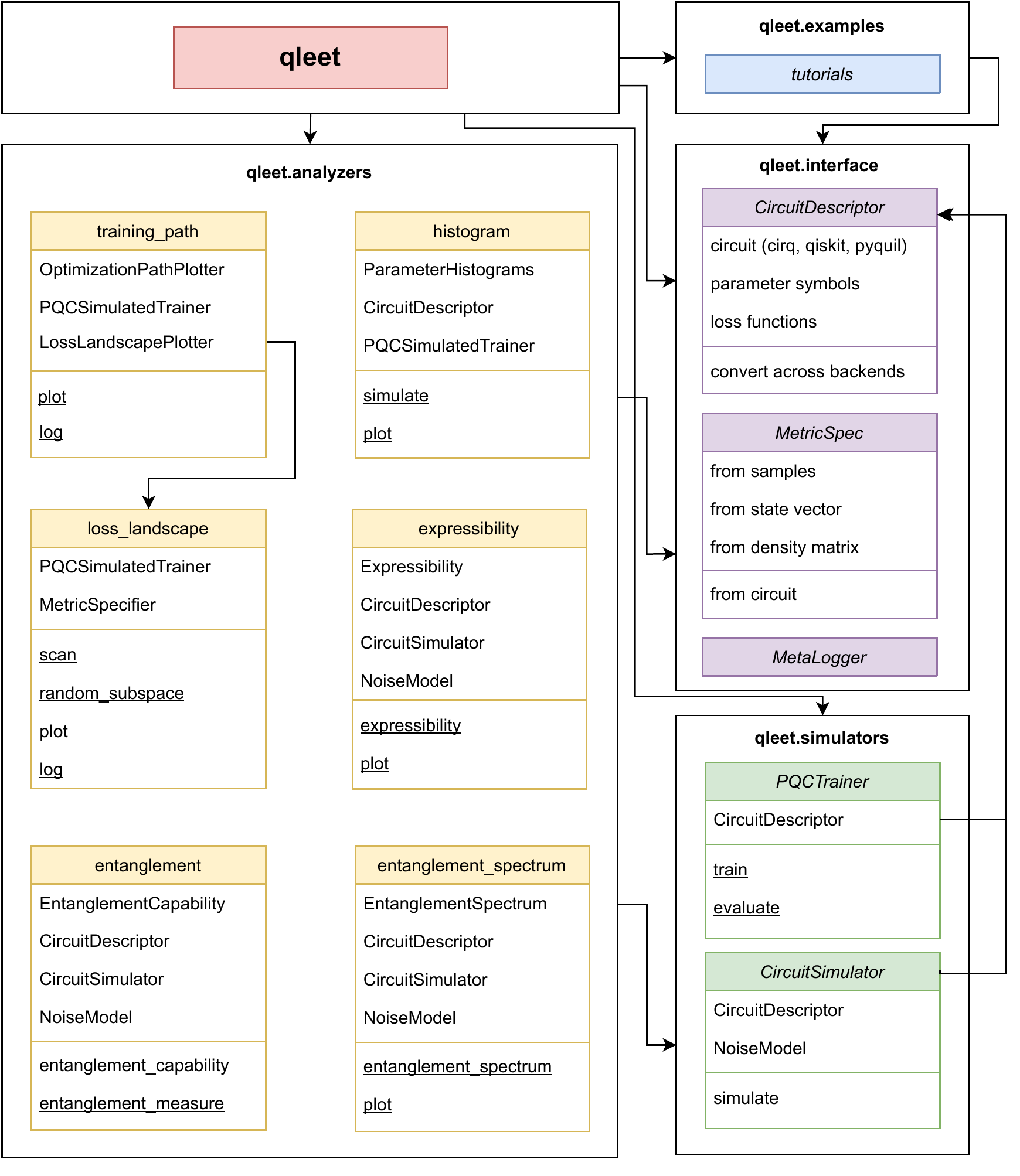}
    \caption{The architecture stack for qLEET: Each block directed from qLEET represents a module. For the analyzers and simulators modules, each sub-block represents a submodule with class objects defined and used them (camel case) and function methods provided by them (underlined). For the interface module, each block represents the class objects defined in it (camel case header) and contains succinct description of their inputs and outputs.}
    \label{fig:qleet-architecture}
\end{figure*}

In this work, we present a python library called qLEET \footnote{\href{https://github.com/QLemma/qleet}{https://github.com/QLemma/qleet}}, \cite{qleet-zenodo}. The primary motivation behind the development of qLEET stems from this need to have a framework for analyzing the capabilities of parameterized quantum circuits and comparing their performances. It does so by allowing users to study various properties related to the behavior of PQCs and assess their effectiveness for a given problem instance. In particular, it will enable visualization of the loss landscape of a PQC for a given objective function and its training trajectory in the parameter space. Furthermore, it allows the calculation of some essential properties of PQCs, such as their expressibility and entangling power \cite{10.1002/qute.201900070}. It is integrable with other popular libraries such as Qiskit \cite{comp_qiskit}, Cirq \cite{comp_cirq}, or PyQuil \cite{ccquad_Pyquil} and also supports instruction-set languages like OpenQASM \cite{2021arXiv210414722C} and Quil \cite{ccquad_Pyquil}.

\textit{Structure} - In Sec. \ref{sec:overview} we present an overview of the architecture stack of qLEET. Then in Sec. \ref{sec:training} and Sec. \ref{sec:challenges}, we demonstrate the use of qLEET in the context of analyzing training of PQCs and mitigating the challenges associated with them. Finally, in Sec. \ref{sec:conclusion}, we conclude with a discussion about our current limitation and possible future extensions of this work.

\section{\label{sec:overview}Overview}

All the functionalities present in qLEET are grouped under four modules, which reside under the top-level module called \texttt{qleet}. Each such module provides modularity in feature development and interacts with one another via a specified workflow or API. We present the complete architecture stack for qLEET in Fig. \ref{fig:qleet-architecture}, listing down the following modules and identifying the interactions within them:

\begin{enumerate}

	\item \textbf{Interface module}: \texttt{qleet.interface} serves as the interface for users to build workflow of the variational computation by specifying the parameterized quantum circuit (PQC) along with its key components like symbolic placeholders for variational parameters ($\vec{\theta}$), an objective or a cost function ($\mathcal{C}$) as an observable in Pauli basis and some metrics for evolving the circuit to the final state defined by \texttt{MetricSpec}. It also contains \texttt{CircuitDescriptor}, which allows for the building of PQC using any supported framework, therefore making the computation software agnostic, and \texttt{MetaLogger}, which maintains the record for events that happen during qLEET's execution. 

	\item \textbf{Simulators module}: \texttt{qleet.simulator} contains the  simulation engine for performing the computation. Depending upon the type of workflow you want to execute, you can choose between \texttt{PQCTrainer} and \texttt{CircuitSimulator} for running training routing and for performing standalone circuit simulation, respectively. At this stage, you may also describe the simulation environment for the computation by providing a noise model for the system. 

	\item \textbf{Analyzers module}: \texttt{qleet.analyzers} performs execution of \texttt{CircuitDescriptor} object using \texttt{PQCTrainer} or \texttt{CircuitSimulator} functions present in the \texttt{qleet.simulator} module. Therefore, \texttt{qleet.analyzers} acts as a linkage between the previous two modules and is responsible for estimating various essential properties regarding PQC. These include loss landscape and training trajectory calculation or histogram prediction for variational computation and expressibility, entangling power and entanglement spectrum calculations for a given ansatz structure. This module also offers plotting functionality for some of these features.
	
	\item \textbf{Example module}: \texttt{qleet.examples} contains basic set of introductory tutorials and predefined templates for users to get started with using qLEET and contribute to it. These include examples of using \texttt{qleet.analyzers} for various kinds of calculations, as mentioned before.

\end{enumerate} 

We maintain the consistency of our codebase via unit testing, type checking, and format checker via pytest \cite{pytestx.y}, mypy \cite{mypy}, and black \cite{black}, respectively. Overall, we aim for the architecture stack for qLEET to follow object-oriented design principles, which helps us create a clean and modular software tool that is easy to test, debug, and maintain in the future. 


\section{\label{sec:features}Features}

This section presents the theory and examples for the features supported by the qLEET. We begin by introducing the idea of the trainability of a parameterized quantum circuit (PQC). From there, we would motivate the idea of studying different properties related to PQC to improve and analyze its trainability. We end the discussion in each subsection by demonstrating how modules in \texttt{qleet} can be used for analyzing the mentioned properties. 

\subsection{\label{sec:training}Trainability of PQCs}

We consider an N-qubit PQC $\hat{U}(\vec{\theta})$ with an objective function defined by a Hermitian observable $O$ in the Pauli basis. For an input quantum state $\rho$, the process of training is defined as minimizing the following function $\mathcal{C}$:
\begin{equation}
	\min \mathcal{C}(\vec{\theta}) = \min \text{Tr}[O \hat{U}(\vec{\theta}) \rho \hat{U}^{\dagger}(\vec{\theta})] = \min \text{Tr}[O \rho(\vec{\theta}) ]
\end{equation}
A PQC $\hat{U}(\vec{\theta})$ evolves the input state $\rho$ to a parameterized target state $\rho(\vec{\theta})$ and to minimize $C(\vec{\theta})$ we update paramters $\vec{\theta}$ via some classical optimization routine such that:
\begin{equation}
	\vec{\theta}^{\,k+1} = \vec{\theta}^{\,k} - \gamma f(\nabla_{\vec{\theta}})\ \mathcal{C}(\vec{\theta}^{\,k}), \quad f(\textbf{0})= 0 
\end{equation}
\begin{figure}[!tp]
    \centering
    \includegraphics[width=0.6\linewidth]{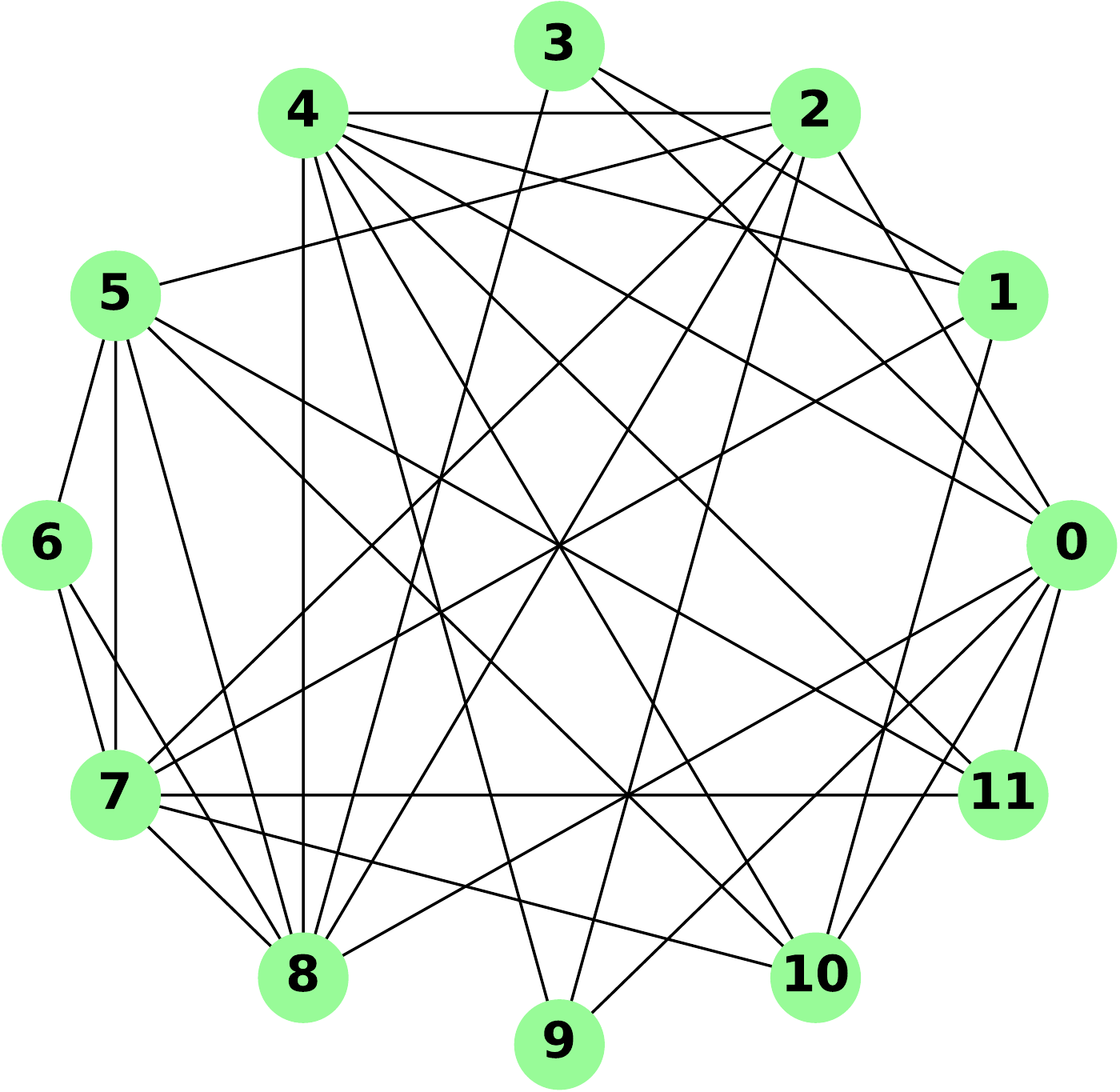}
    \caption[Problem graph for QAOA]{Problem graph considered for MaxCut using QAOA. It is generated as an Erdos-Renyi graph with $12$ nodes and $0.5$ edge probability.}
    \label{fig:qoao-maxcut-graph}
\end{figure}
Therefore, for successfully training a PQC, we would require contributions from any variational parameter $\theta_v$ to $\nabla_{\vec{\theta}}$, i.e., $\partial\mathcal{C}/\partial\theta_v$ to be non-vanishing, non-exploding and unbiased. This means that we expect $\mathbb{E}(\partial\mathcal{C}/\partial\theta_v) = 0$ and $\text{Var}(\partial\mathcal{C}/\partial\theta_v) > 0$  $\forall \theta_v \in \vec{\theta}$. However, this is not always the case, as we would see later in Sec. \ref{sec:challenges}. To better understand this behaviour, it is critical to look at the evolution of $\mathcal{C}$ with respect to changes in variational parameters for which computing loss landscape and training path is beneficial. Furthermore, it has also been shown that circuits with large expressibility seem to have vanishing gradients, i.e., $\nabla_{\vec{\theta}}\ \mathcal{C} \rightarrow 0$. Hence, it is also crucial to not just look at the evolution of $\mathcal{C}$ but also get insights from the intrinsic properties of the PQC itself, such as its expressibility and entangling power.

\begin{figure*}[htp]
    \centering
    \begin{subfigure}[b]{0.32\linewidth}
        \includegraphics[width=\textwidth]{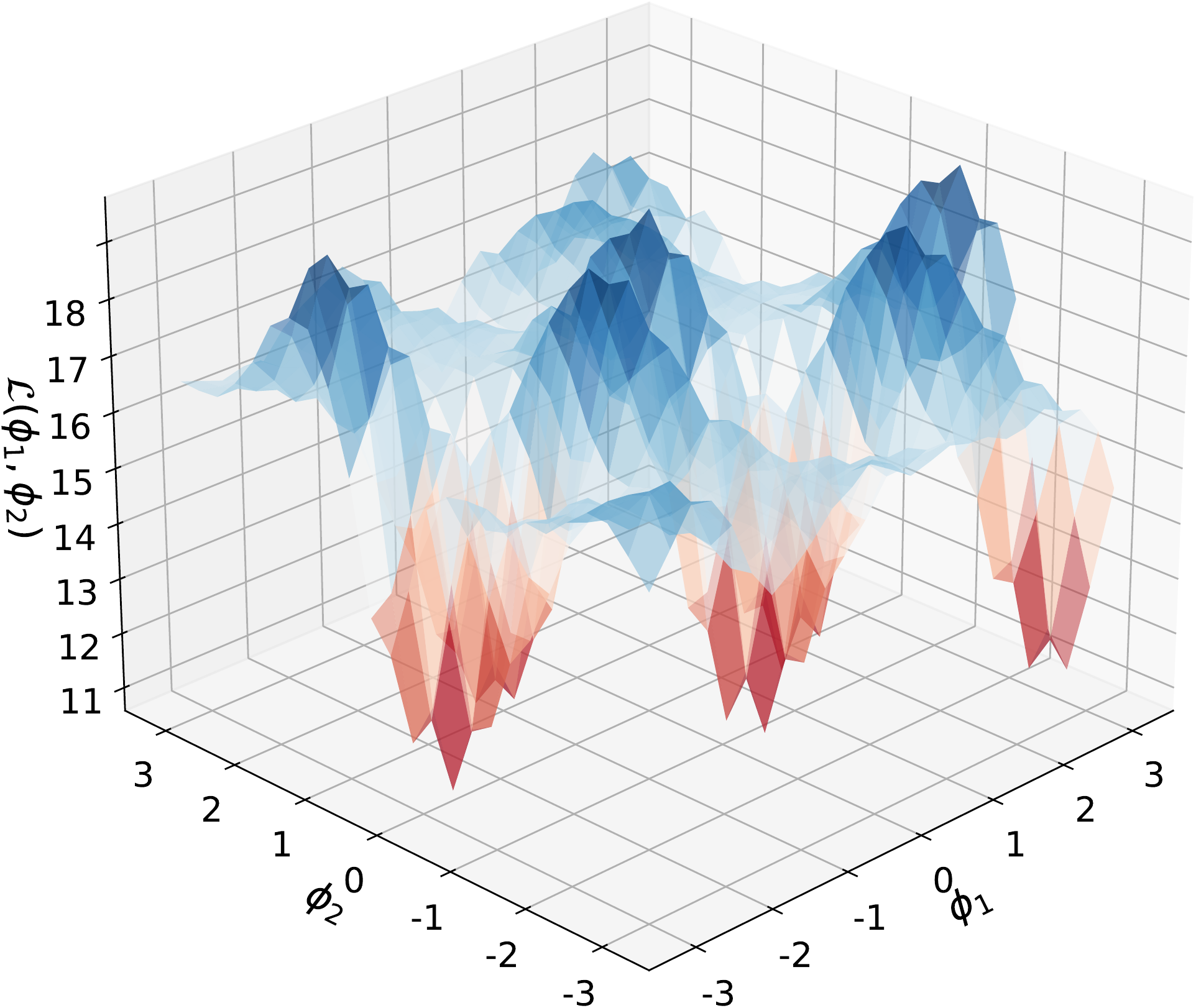}
        \caption{Loss Landscape for $p=1$\label{fig:loss-p1}}
    \end{subfigure}
    \begin{subfigure}[b]{0.32\linewidth}
        \includegraphics[width=\textwidth]{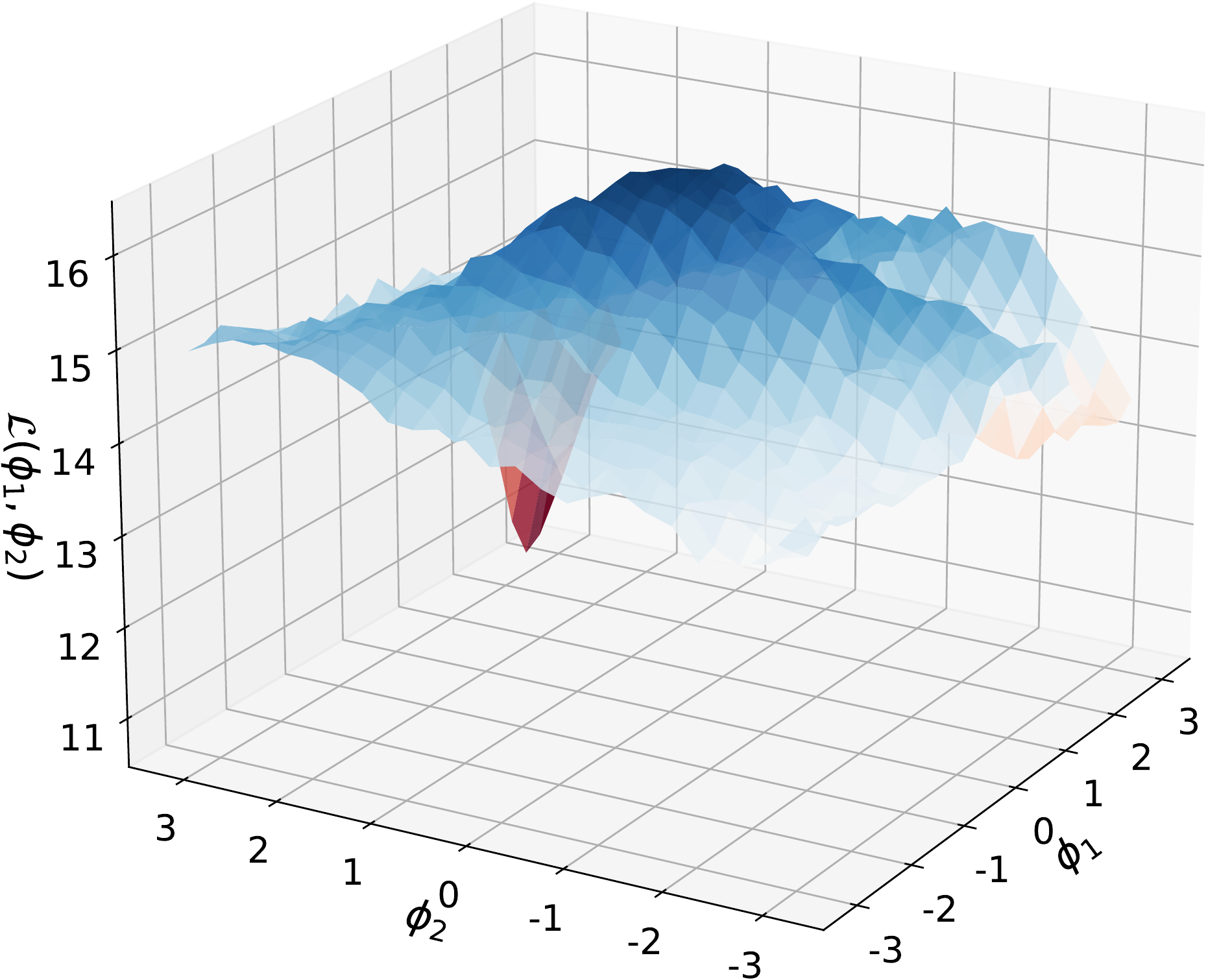}
        \caption{Loss Landscape for $p=4$\label{fig:loss-p4}}
    \end{subfigure}
    \begin{subfigure}[b]{0.32\linewidth}
        \includegraphics[width=\textwidth]{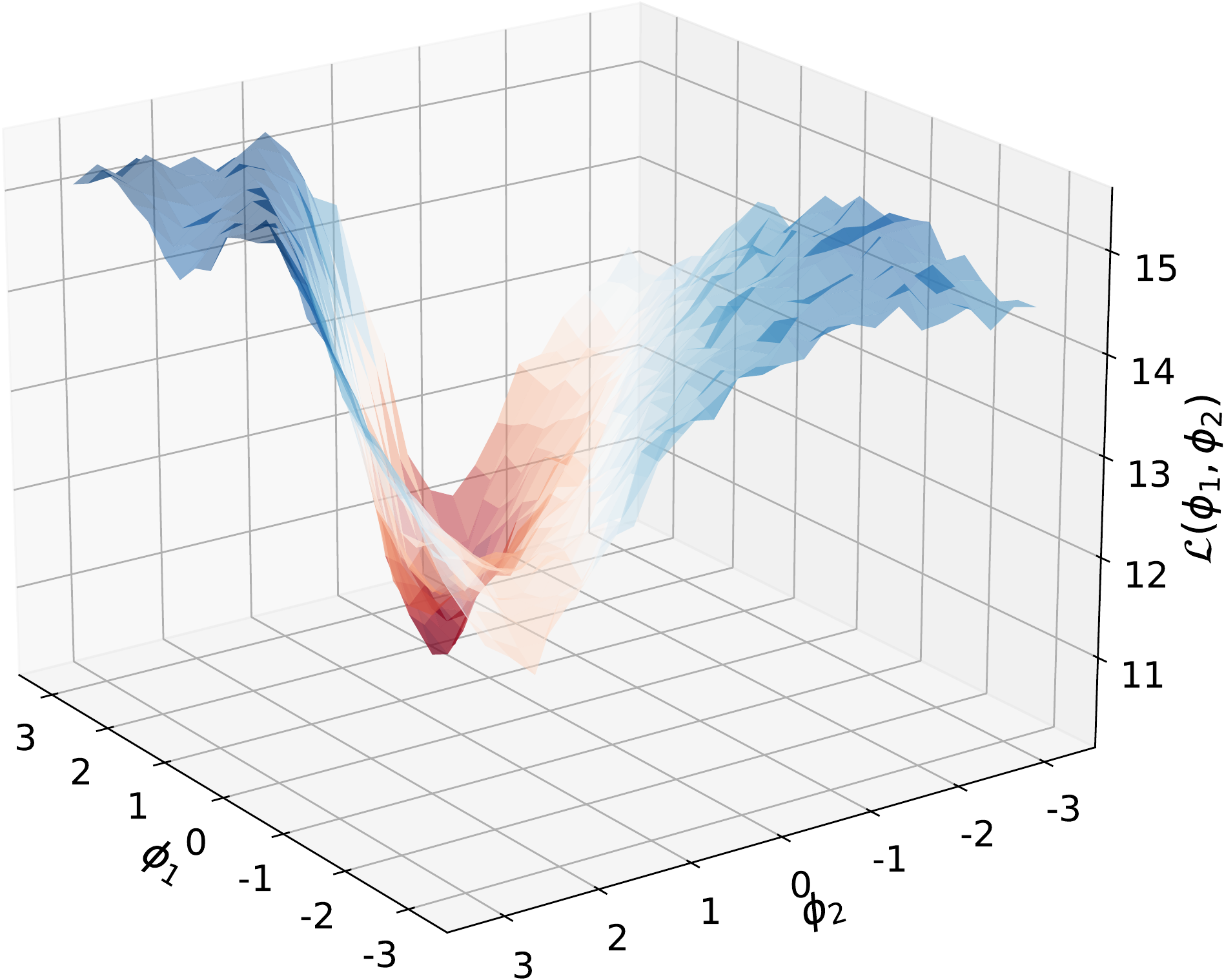}
        \caption{Loss Landscape for $p=8$\label{fig:loss-p8}}
    \end{subfigure}%
    \hfill\newline
    \begin{subfigure}[b]{0.32\linewidth}
        \includegraphics[width=\textwidth]{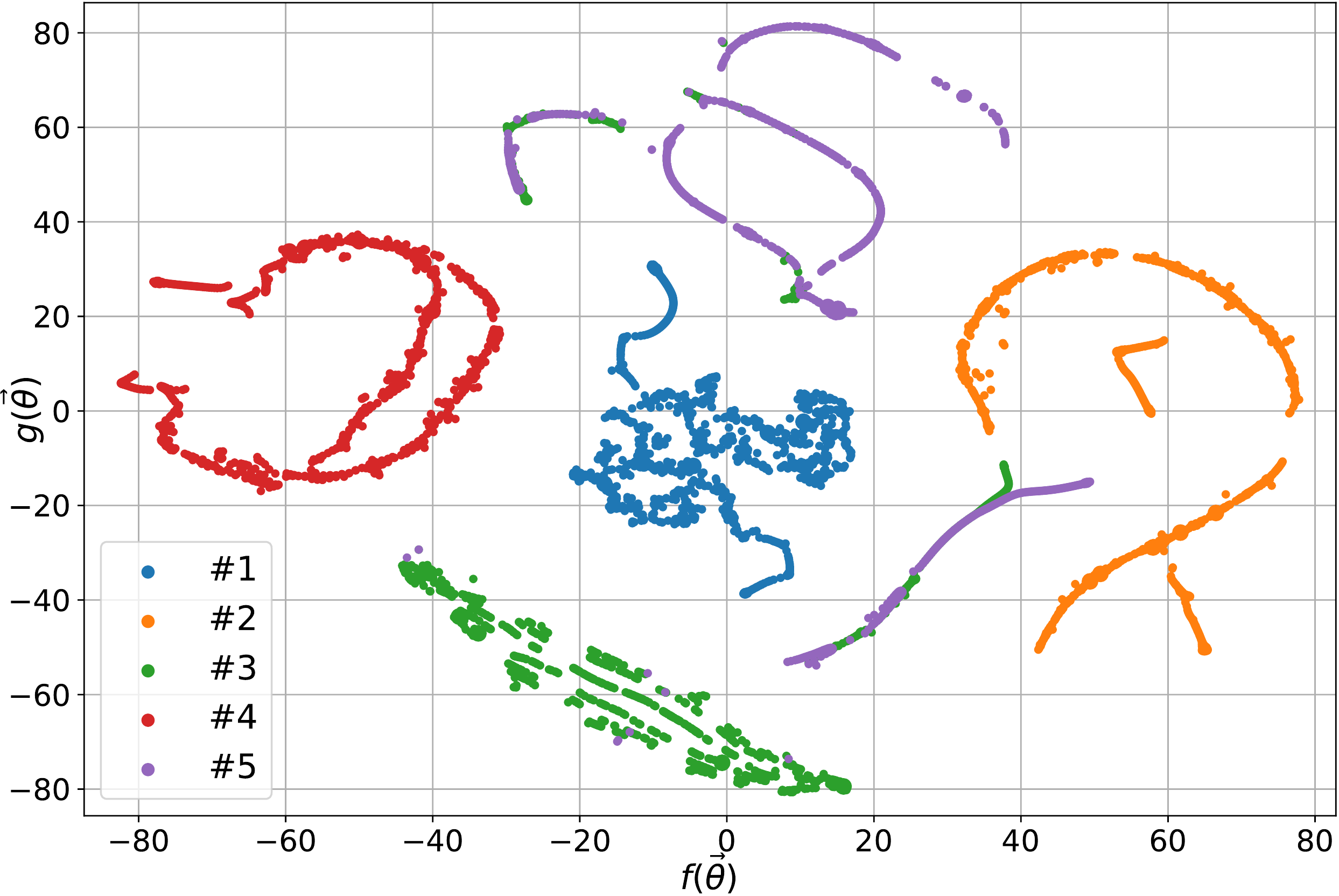}
        \caption{Training trajectories for $p=1$\label{fig:train-p1}}
    \end{subfigure}
    \begin{subfigure}[b]{0.32\linewidth}
        \includegraphics[width=\textwidth]{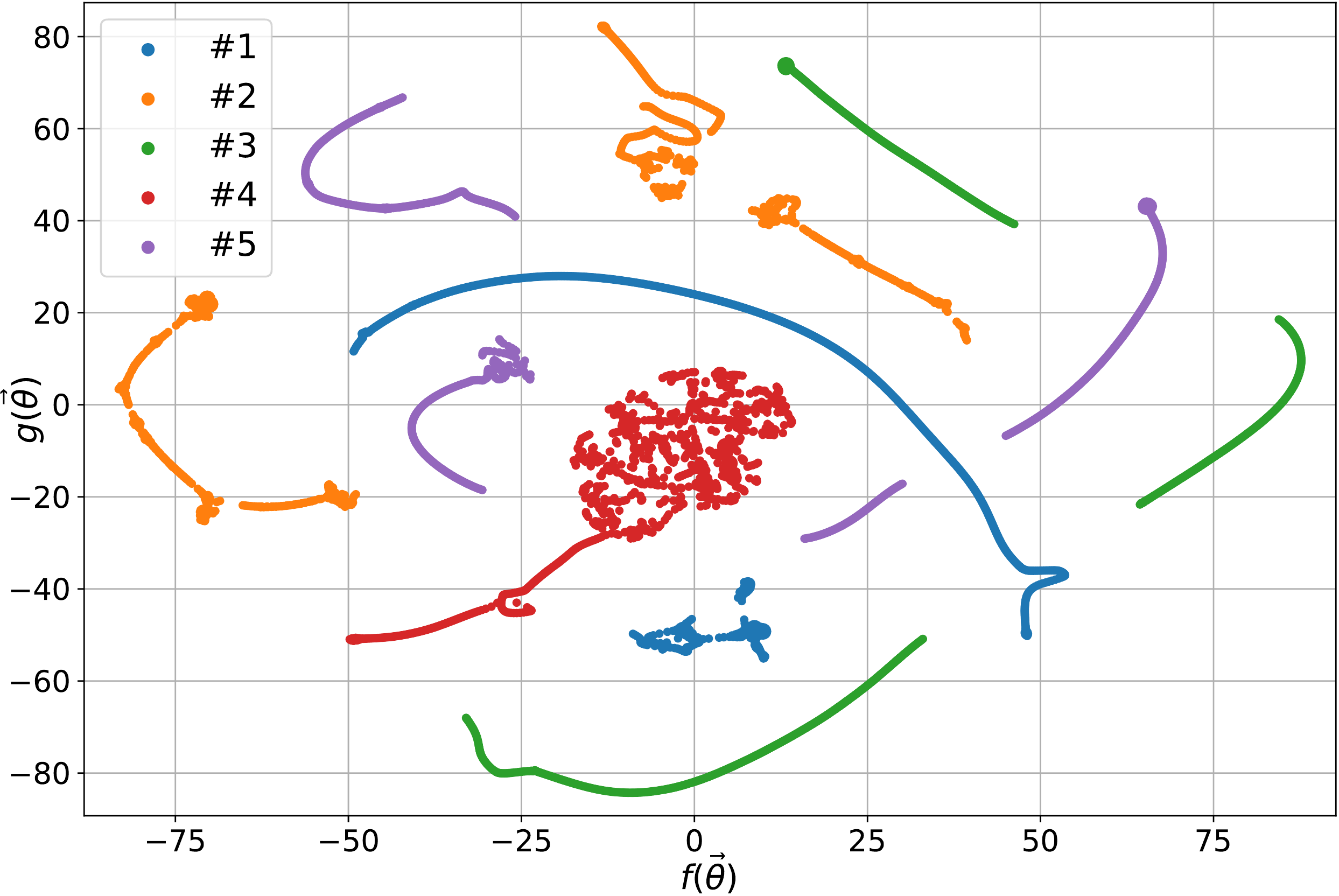}
        \caption{Training trajectories for $p=4$\label{fig:train-p4}}
    \end{subfigure}
    \begin{subfigure}[b]{0.32\linewidth}
        \includegraphics[width=\textwidth]{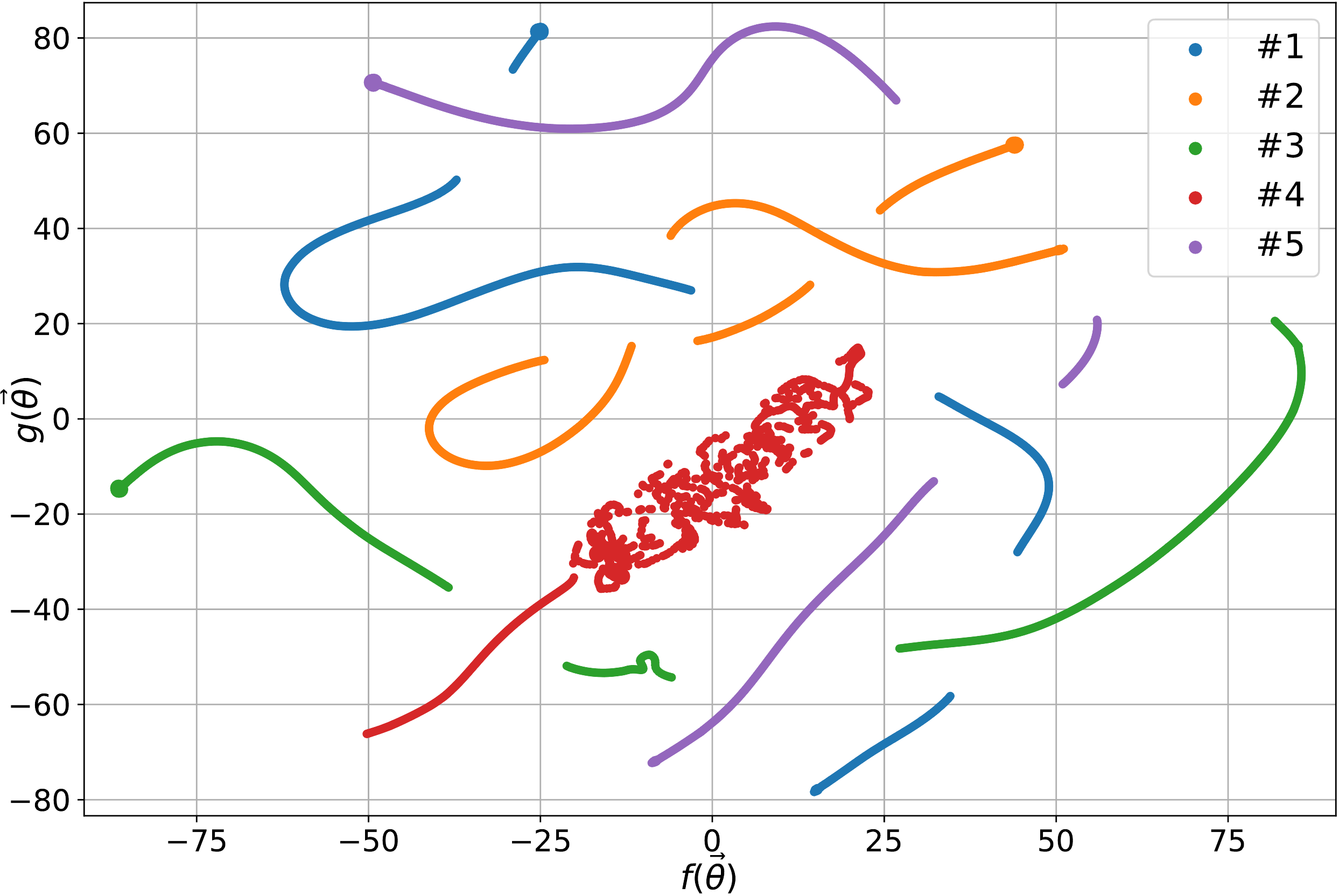}
        \caption{Training trajectories for $p=8$\label{fig:train-p8}}
    \end{subfigure}%
    \caption{Loss landscape and training trajectories plots for solving the MaxCut problem using QAOA routine implemented with qLEET for the graph presented in Fig. \ref{fig:qoao-maxcut-graph}. The training trajectories have been plotted for five instances of training with different random initializations of variational parameters $\vec{\theta}$ for each value of $p\in\{1, 4, 8\}$, where $p$ denotes the number of times QAOA ansatz is repeated and functions $f(\vec{\theta})$ and $g(\vec{\theta})$ represents non-linear functions obtained after dimensionality reduction using t-SNE}
    \label{fig:loss-land-train-traj}
\end{figure*}

\subsection{Loss Landscape}

Loss landscape is a visual representation of the loss values or the $\mathcal{C}(\vec{\theta})$ around the trainable variational parameter space of the PQC. This inspection is usually done around the optimal variational parameters $\vec{\theta}^{*}$ to identify features like local minima, ridges, and valleys present in the loss surface. Such analysis helps in analyzing smoothness off the surface, indicating the ease with which a gradient-based optimizer might be able to perform on it \citep{loss-landscapes, 2021arXiv211104695R}. 

\begin{figure*}[!tp]
    \centering
    \includegraphics[width=0.82\textwidth]{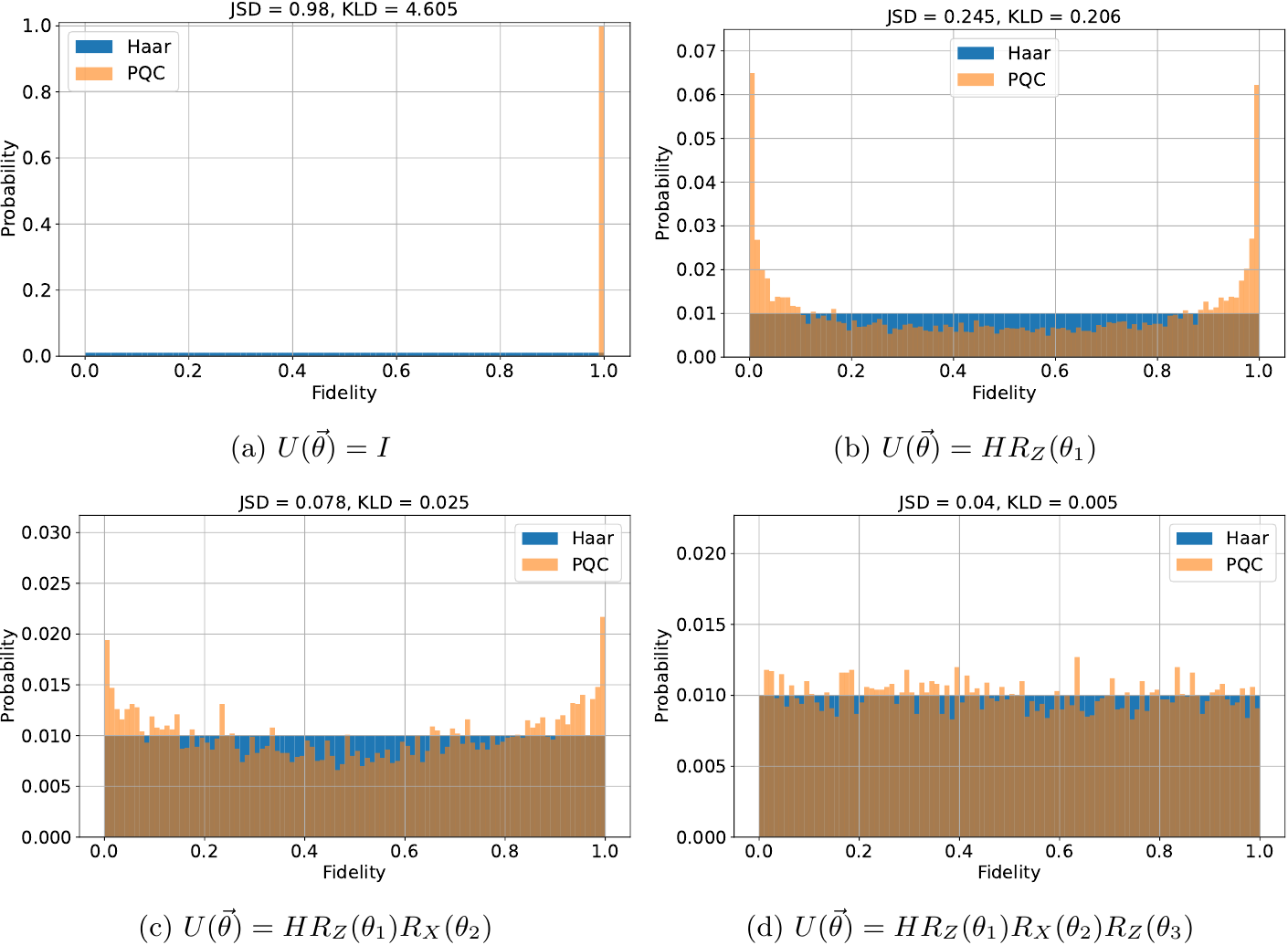}
    \caption[Quantifying expressibility for single-qubit circuits]{Quantifying expressibility for single-qubit circuits. For each of the four circuits show here, 1000 sample pairs of circuit parameter vectors were uniformly drawn, corresponding to 2000 parameterized states. Histograms of estimated fidelities (orange) are shown, overlaid with fidelities of the Haar-distributed ensemble (blue), with the computed Kullback-Leibler (KL) divergence and Jensen-Shannon Distance (JSD) reported above the histograms.}
    \label{fig:expressibility}
\end{figure*}

For example, in Fig. \ref{fig:loss-land-train-traj}, we look at the loss landscape associated with solving the MaxCut problem using the QAOA algorithm \cite{2014arXiv1411.4028F} for an Erdos-Renyi graph (Fig. \ref{fig:qoao-maxcut-graph}). We see that as the number of layers of QAOA ansatz, parameterized by $p$, are increased, the loss landscape becomes much smoother, and local minima pits disappear. Therefore, it would be much easier for a descent-based optimizer to traverse to global minima in case of higher $p$. This and similar loss landscape calculations in qLEET are done using the \texttt{loss\_landscape} function present in the analyzer module. As shown in Eq. \ref{eqn:loss-landscape-plot}, we compute the value of the loss function $\mathcal{L}$ for all the coordinates $\vec{\phi}\ (=\mathbf{\Phi}^T\, \vec{\theta}\,$) in an orthonormalized 2-D subspace $S$ with basis vectors $\vec{\Phi}_i$ sampled from the whole trainable variational parameter space and origin corresponding to the optimal variational parameters $\theta^*$.
\begin{equation}\label{eqn:loss-landscape-plot}
    \begin{split}
        \mathcal{L}(\phi_0, \phi_1) 
        &= \mathcal{C}(\vec{\theta}^* + \vec{\theta}^\prime\,), \quad \theta^\prime_i = (\mathbf{\Phi}\, \vec{\phi}\, )^{\,T} \\ 
        &= \sum_{O} \text{Tr}\Bigg[O\rho \bigg(\vec{\theta}^* + \vec{\theta}^\prime\, \bigg) \Bigg]
    \end{split}
\end{equation}
We gather different information about the loss of landscape based on how we choose to perform the sampling. For example, using principal component analysis (PCA) over the set of variational parameters $\vec{\theta}$ at each training step would give us the vectors $\vec{\Phi}_i$ that represent the directions in parameter space for which major changes happen during that training step. Similarly, other methods for obtaining subspace could be used, such as doing random sampling of basis vectors or t-SNE (t-Distributed Stochastic Neighbor Embedding) of the parameter vectors encountered in the training trajectory. All such methods provide beneficial insights about the structure of the loss landscape using which one could adapt their training strategy by tweaking the optimization routine, evaluation metric, etc. 

\subsection{Training Trajectory}

In many cases, just looking at the loss landscape for a given PQC model is not enough as we define the subspace $S$ using two of many possible directions as axes by taking linear combinations of variational parameters, while the loss landscape itself is highly nonlinear. Moreover, the high dimensionality of the parameter space makes the task of visualization of loss landscape extremely challenging. However, both of these difficulties can somewhat be alleviated by visualizing the loss landscape via the evolution of variational parameters of PQC during training in low dimensions. This evolution of variational parameters can be realized as the training trajectory for the PQC, and plotting them over several re-initializations helps us learn about the convergence properties of the PQCs and their optimization schedules. 

In qLEET, training trajectories are calculated inside the analyzer module by the \texttt{training\_path} function. We use the entire set of variational parameters $\vec{\theta}^{t}$ to generate the trajectory over all re-initialization for every time step $t$ in the training process. We project the parameter vectors down to an orthonormalized 2-D subspace $S$ using techniques such as PCA \cite{Jolliffe2016}, t-SNE \cite{NIPS2002_6150ccc6}, or PHATE \cite{Moon2019}. Similar to the case of loss landscape visualization, each of the mentioned techniques reveals different trajectory characteristics depending on its ability to preserve both global and local structures of higher-dimensional data in low dimensional subspace. Furthermore, the 2-D projections of the parameter trajectories can also be plotted on the loss surface, with the loss values as its third axis \citep{training-trajectories}.

For example, we present the training trajectories with t-SNE projection in Fig. \ref{fig:loss-land-train-traj} for the same MaxCut problem that we discuss in the previous subsection about the loss landscape. We look at five different training instances for each $p$, where we begin with randomized initialization of variational parameters $\vec{\theta}$ every time. We see that for $p=1$ evolutions of $\vec{\theta}$ for every instance happen in their own respective clusters, suggesting the optimizer unsuccessfully gets stuck for different local minima every time. In contrast, for both $p=4$ and $p=8$, we see much lesser clusters formation and more intercrossing, hinting at certain parameters $\theta_k$ evolving to the same values while the optimizer reaches the global minima. 


\subsection{Expressibility}

\begin{figure}[!t]
    \centering
    \includegraphics[width=\linewidth]{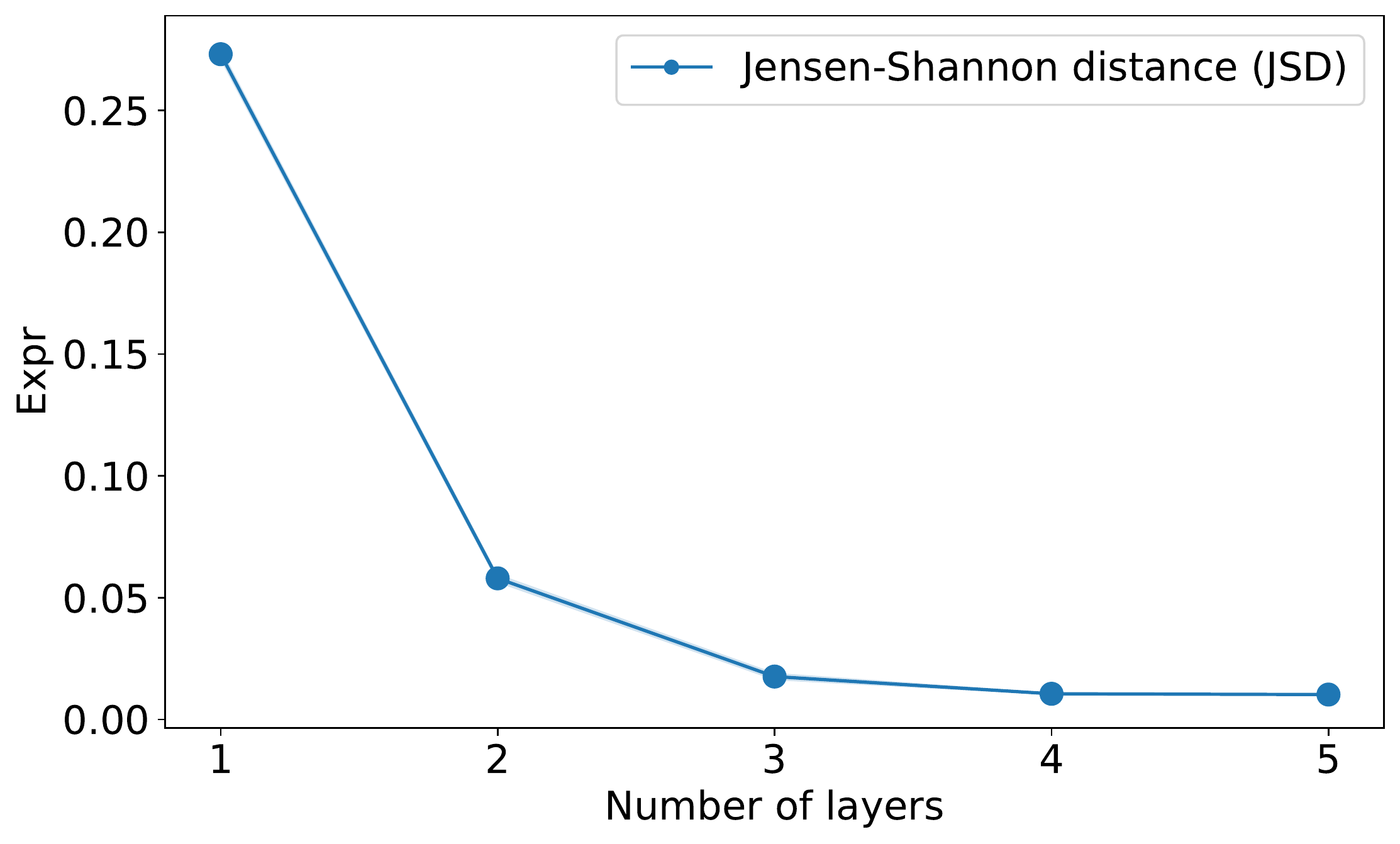}
    \caption[Visualizing entanglement spectrum for parameterized quantum circuits]{Measuring expressibility for the parameterized quantum circuit $U(\vec{\theta}) =  \prod_{1}^{L}\big(\bigotimes_{i=1}^{5}R_x(\theta_i^1)R_z(\theta_i^2)R_x(\theta_i^3) \ldots \bigotimes_{i<j}CX(i, j)\big)$ using the Jensen-Shannon distance (JSD) measure as a function of number of layers $L$. }
    \label{fig:expressibility-measure}
\end{figure}

We generate a distribution of states $\rho(\vec{\theta})$ for a PQC $\hat{U}(\vec{\theta})$ by randomly sampling over the variational parameter space. We quantify the deviation of this distribution from the one obtained from the maximally expressive Haar distribution as the \textit{Expressibility} of the given ansatz.
\begin{equation}\label{qleet:eq3}
    A^{(t)} =\left\Vert \int_\text{Haar}\rho^{\otimes t} \text{d}\rho - \int_{\vec{\theta}}\rho(\vec{\theta})^{\otimes t} \text{d}\rho(\vec{\theta}) \right\Vert_\text{HS}^2\,
\end{equation}
where $\int_\text{Haar}\text{d}\rho$ denotes the integration over the states $\rho$ produced by the unitaries sampled according to the Haar measure over the unitary group $\mathcal{U}$, $t$ represent the $t^{\text{th}}$ moment,  and $\left\Vert A \right\Vert_\text{HS}^2$ is the Hilbert-Schmidt norm calculated as $\text{Tr}(A^\dagger A)$. We compute Eq. \ref{qleet:eq3} as the divergence between the distribution of fidelities $\mathcal{F}(\rho, \sigma) = \left(\text{Tr}\sqrt{\sqrt{\rho}\sigma\sqrt{\rho}} \right)^2$ \cite{Jozsa1994} of the states $(\rho, \sigma)$ obtained from the unitaries $U_{\rho}, U_{\sigma} \in \mathcal{U}_{\text{PQC}}$ (or $ \mathcal{U}_{\text{Haar}}$), where $\mathcal{U}_{\text{PQC}}$ is the ensemble of parameterized unitaries describing the ansatz for uniformly sampled $\vec{\theta}$ and $\mathcal{U}_{\text{Haar}}$ is the ensemble of Haar random unitaries \cite{10.1002/qute.201900070}.

\begin{equation}
    \text{Expr} = D(\hat{P}_{PQC}(\mathcal{F}; \vec{\theta}) | P_{Haar}(\mathcal{F})), \quad \text{Expr} \geq 0
\end{equation}
According to this definition, a PQC $U(\vec{\theta})$ is more expressible if the distribution of state fidelities generated by the ansatz circuit $U(\vec{\theta})$ is closer to the one generated by the unitaries $U_{\text{Haar}}$ sampled uniformly from the unitary group $\mathcal{U}$. Therefore, the smaller the \textit{Expr} value, the more is the expressibility of the parameterized unitary. We see this in  Fig. \ref{fig:expressibility}, where we compare the fidelity distribution of PQC and Haar random states with respect to the number of Pauli rotation gates present in the single-qubit circuits and calculate the \textit{Expr} values for both Kullback-Leibler (KL) and Jensen-Shannon (JS) divergence. Furthermore, in Fig. \ref{fig:expressibility-measure}, we measure the increasing expressibility of the five qubit ansatz $U(\vec{\theta}) =  \prod_{1}^{L}\big(\bigotimes_{i=1}^{5}R_x(\theta_i^1)R_z(\theta_i^2)R_x(\theta_i^3) \ldots \bigotimes_{i<j}CX(i, j)\big)$, where we see how expressibility increases with the number of layers $L$.  Finally, we note that, in addition to experiments like these, \texttt{expressibility} function in qLEET can also be used to predict the likelihood of whether the given PQC would be able to represent an unknown N-qubit target state and do a comparative analysis between different ansätze.

\begin{figure}[!t]
    \centering
    \includegraphics[width=\linewidth]{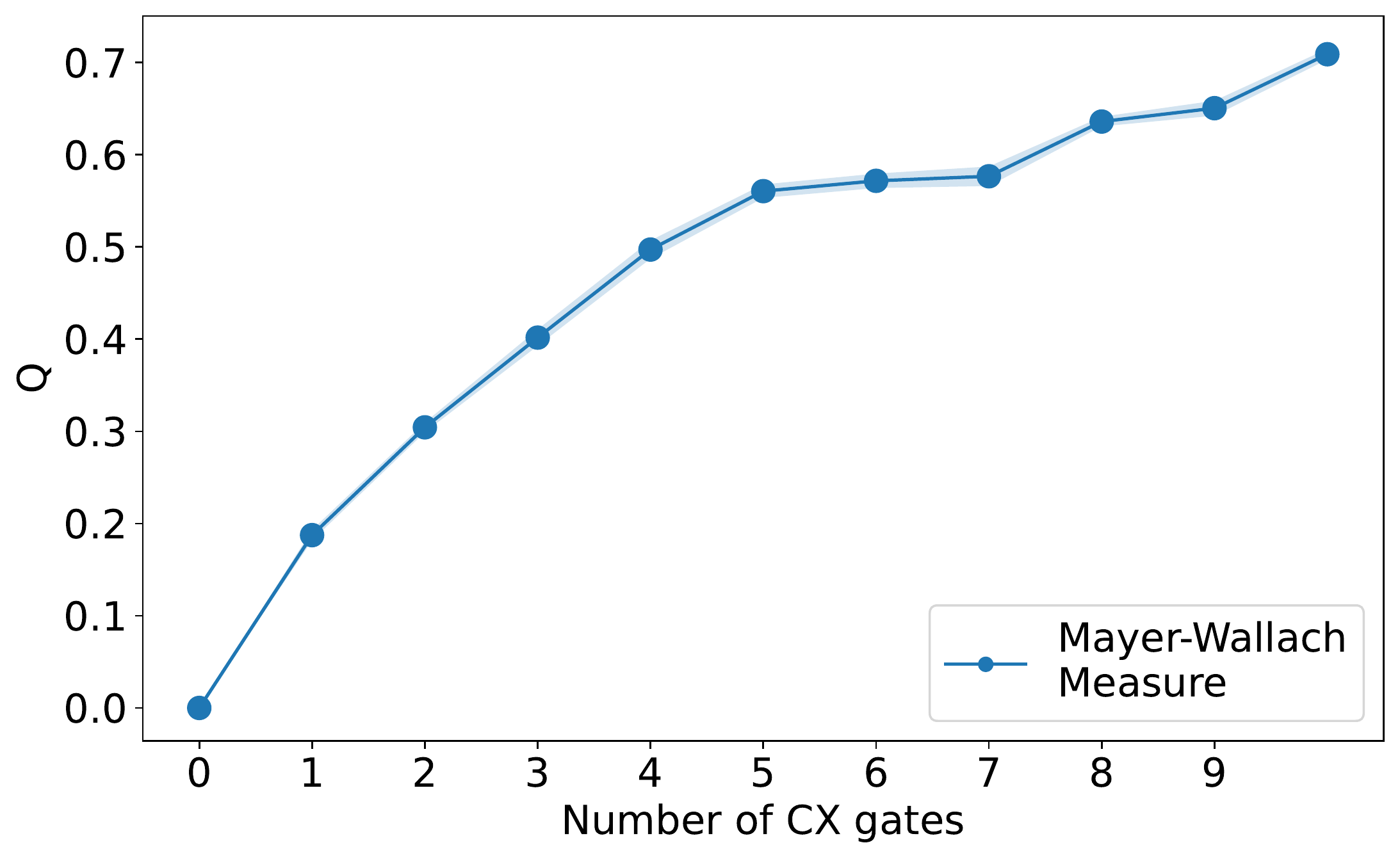}
    \caption[Visualizing entanglement spectrum for parameterized quantum circuits]{Measuring entangling power for the parameterized quantum circuit $U(\theta)$ using the Mayer-Wallach measure as a function of the number of CX (or CNOT) gates appended to the circuit $U(\vec{\theta}) = \bigotimes_{i=1}^{5}R_x(\theta_i^1)R_z(\theta_i^2)R_x(\theta_i^3)$.}
    \label{fig:entanglement-capability}
\end{figure}

\subsection{Entangling Capability}

A fundamental property that makes quantum computation different from the classical one is the existence of entanglement in the system, which can be potentially exploited to gain a computational advantage. Hence, it is essential to quantify its ability to generate entanglement in the system to assess the effectiveness of a parameterized quantum circuit. We use entanglement measures to capture different properties of multipartite entanglement present in the system. The first measure that we use is the Meyer-Wallach $Q$ measure \cite{10.1002/qute.201900070, doi:10.1063/1.1497700} in which the amount of entangled states produced by a PQC is estimated by measuring the average entanglement between individual qubits and the rest of the system. In this context, the entangling capability of a PQC can be defined directly via the considered entanglement measure $Q$ averaged over all states $\rho(
\vec{\theta})$ generated by the PQC from the uniform sampling of variational parameters $\vec{\theta}$:

\begin{equation}
	Q = \frac{2}{|\vec{\theta}|}\sum_{\vec{\theta}^{\, i}\in \{\vec{\theta}\}}\Bigg(1-\frac{1}{n}\sum_{k=1}^{n}\text{Tr}(\rho_{k}^{2}(\vec{\theta}^{\,  i}))\Bigg),
\end{equation}
where $\rho_k$ is the density matrix for the state of the $k$-th qubit. In a similar spirit, we can use another entanglement measure called Scott Measure \cite{10.1007/s11128-007-0052-7}, which generalizes the Meyer-Wallach measure using $m$ entanglement measures, each of which will measure the average entanglement between blocks of $m$ qubits and the rest of the system. Therefore, as pointed out before, each measure would give access to different properties related to multipartite entanglement, and as $m$ increases, $Q_m$ becomes more sensitive to correlations of an increasingly global nature. Similar to the previous case, the entangling capability of the PQC can be defined by the value of $Q_m$ measures, averaged over uniformly sampled $\vec{\theta}$ too:
\begin{equation}
    \begin{split}
        Q_{m} &= \frac{2^{m}}{(2^{m}-1) |\vec{\theta}|}\sum_{\vec{\theta}^{\,i} \in \{\vec{\theta}\}} \bigg(1 - \\ 
        & \quad \quad \frac{m! (n-m)!)}{n!}\sum_{|S|=m} \text{Tr} (\rho_{S}^2 (\vec{\theta}^{\,i})) \bigg) \\
        m &= 1, \ldots, \lfloor n/2 \rfloor
    \end{split}
\end{equation}

In qLEET, we perform these calculations inside the $\texttt{entanglement}$ function in the analyzer module, where one can choose between both Meyer-Wallach and Scott measures for any PQC loaded as a $\texttt{CircuitDescriptor}$ object. For example, in Fig. \ref{fig:entanglement-capability}, we use it to plot the entangling capability of a five qubit circuit template $U(\vec{\theta}) = \bigotimes_{i=1}^{5}R_x(\theta_i^1)R_z(\theta_i^2)R_x(\theta_i^3)$ against the numbers of CNOT gates appended to circuit in a pair-wise fasion, i.e., CNOT$(i, j)$, where $i < j$ and $i,j <5$. We see that as the number of CNOT gates are increased, the entangling capability improves. We also notice a region of minimal increase between $[5, 7]$, which can be attributed to addition on qubits which were already transitive correlated.

\subsection{Entanglement Spectrum}

In the previous subsection, we quantified the entangling capability of an ansatz using entanglement measures. However, these measures might be insufficient to fully characterize all the properties related to multipartite entanglement \cite{PhysRevLett.115.267206}. This problem can be tackled by making use of the entanglement spectrum  \cite{PRXQuantum.1.020319}, which is defined as the eigenspectrum of the entanglement Hamiltonian $H_{\text{ent}}$:

\begin{equation}
    H_{\text{ent}} = -\log (\rho_A),
\end{equation}

where the $\rho_A = \text{Tr}_B(\rho)$ is the reduced density matrix of the qubit system obtained by the typical bipartition of the $N$ qubit system into subsystems $A$ and $B$ with $k = \lceil N/2 \rceil$ and $N-k$ qubits, respectively. For states sampled from maximally expressive Haar distribution, the eigenvalues $\xi_k$ of $H_{\text{ent}}$ follows the Marchenko-Pastur (MP) distribution \cite{10.1088/1751-8113/40/3/f04}. Therefore, we can quantify both expressibility and entangling power of the PQC by looking at the eigenspectrum of $H_{\text{ent}}^{\text{PQC}}$, calculated from uniformly sampled variational parameters $\vec{\theta}$. 

\begin{figure}[!t]
    \centering
    \includegraphics[width=\linewidth]{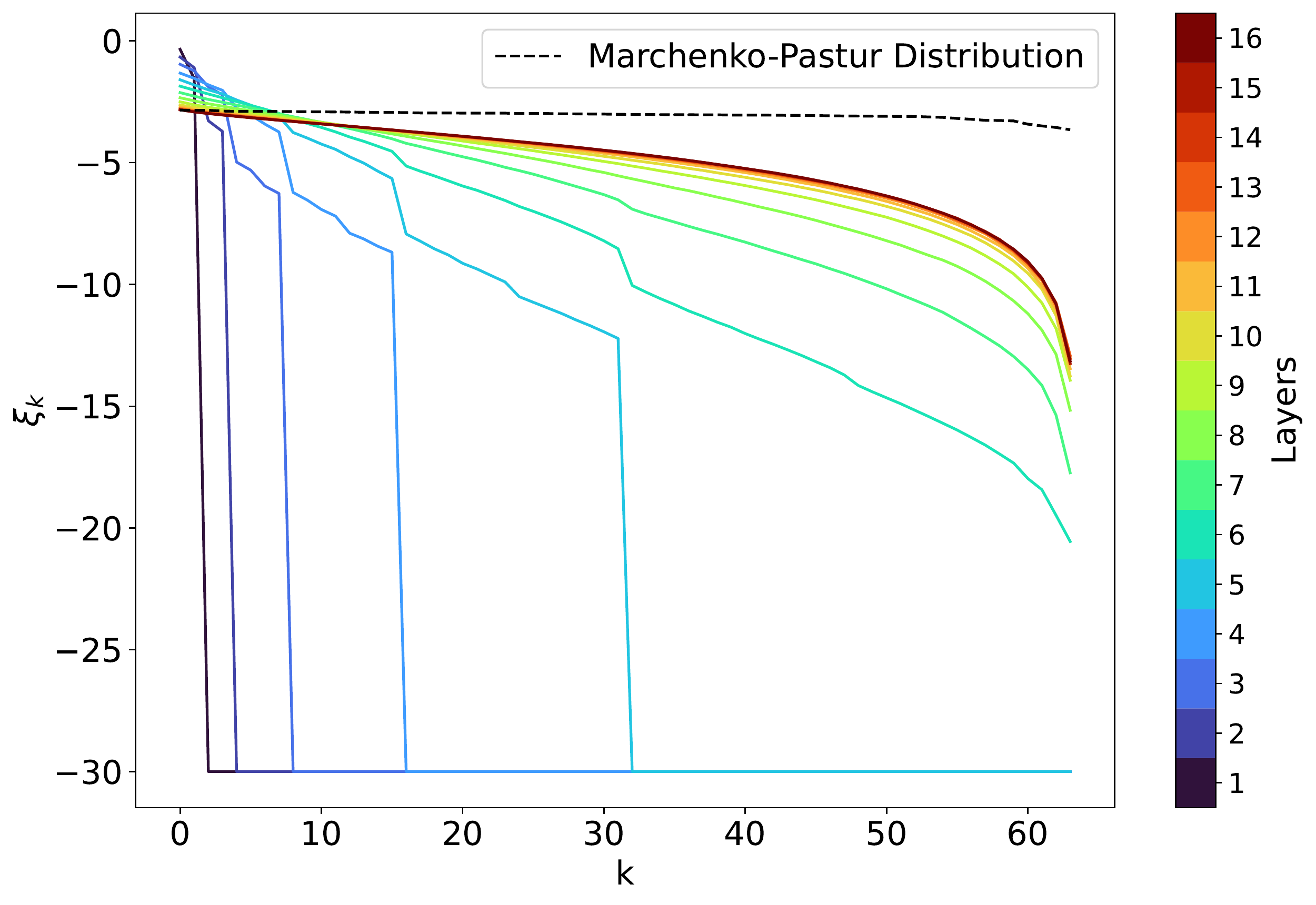}
    \caption[Visualizing entanglement spectrum for parameterized quantum circuits]{Visualizing entanglement spectrum for a PQC $U(\vec{\theta}) = \prod_{1}^{L}\big(\bigotimes_{i=1}^{12}R_x(\theta_i^1)R_z(\theta_i^2)R_x(\theta_i^3) \ldots \bigotimes_{i=1}^{11}CX(i, i+1)\big)$. Here, $\xi_k$ are the eigenvalues of $H_{\text{ent}}^{U(\vec{\theta})}$ arranged in descending order and cut off at $-30$. The solid lines (blue to brown) represents the distribution $\xi_k$ for different layers $L$ and the dotted line (black) represents the ideal Marchenko-Pastur (MP) distribution. We see that as the number of layers is increased, the distribution of $\xi_k$ becomes more similar to MP distribution.}
    \label{fig:entanglement-spectrum}
\end{figure}

In qLEET, \texttt{entanglement\_spectrum} function in the analyzers module can be used for computing and plotting the entanglement spectrum for any given PQC $U(\vec{\theta})$. For example, in Fig. \ref{fig:entanglement-spectrum}, we use it to perform the entanglement spectrum analysis on a 16 qubit PQC, which is made of $L$ layers comprising three rotation gates on each qubit and CNOT gates between adjacent qubits, i.e., $U(\vec{\theta}) = \prod_{l}^{L}\big(\bigotimes_{i=0}^{15}R_x(\theta_i^1)R_z(\theta_i^2)R_x(\theta_i^3)\bigotimes_{i=0}^{14}CX(i, i+1)\big)$. We see that as the number of layers are increased in the ansatz, the eigenvalue distribution becomes more and more closer to the MP distribution. In fact, computing a divergence measure between these two distributions can also be used as a quantification of capability of the ansatz. 

\begin{figure*}[t]
    \centering
    \begin{subfigure}[b]{0.48\textwidth}
    \begin{minipage}
    {.03\textwidth}
        \caption{}
        \label{fig:barren-plateau-1}
    \end{minipage}%
    \begin{minipage}{0.90\textwidth}
        \includegraphics[width=.9\textwidth]{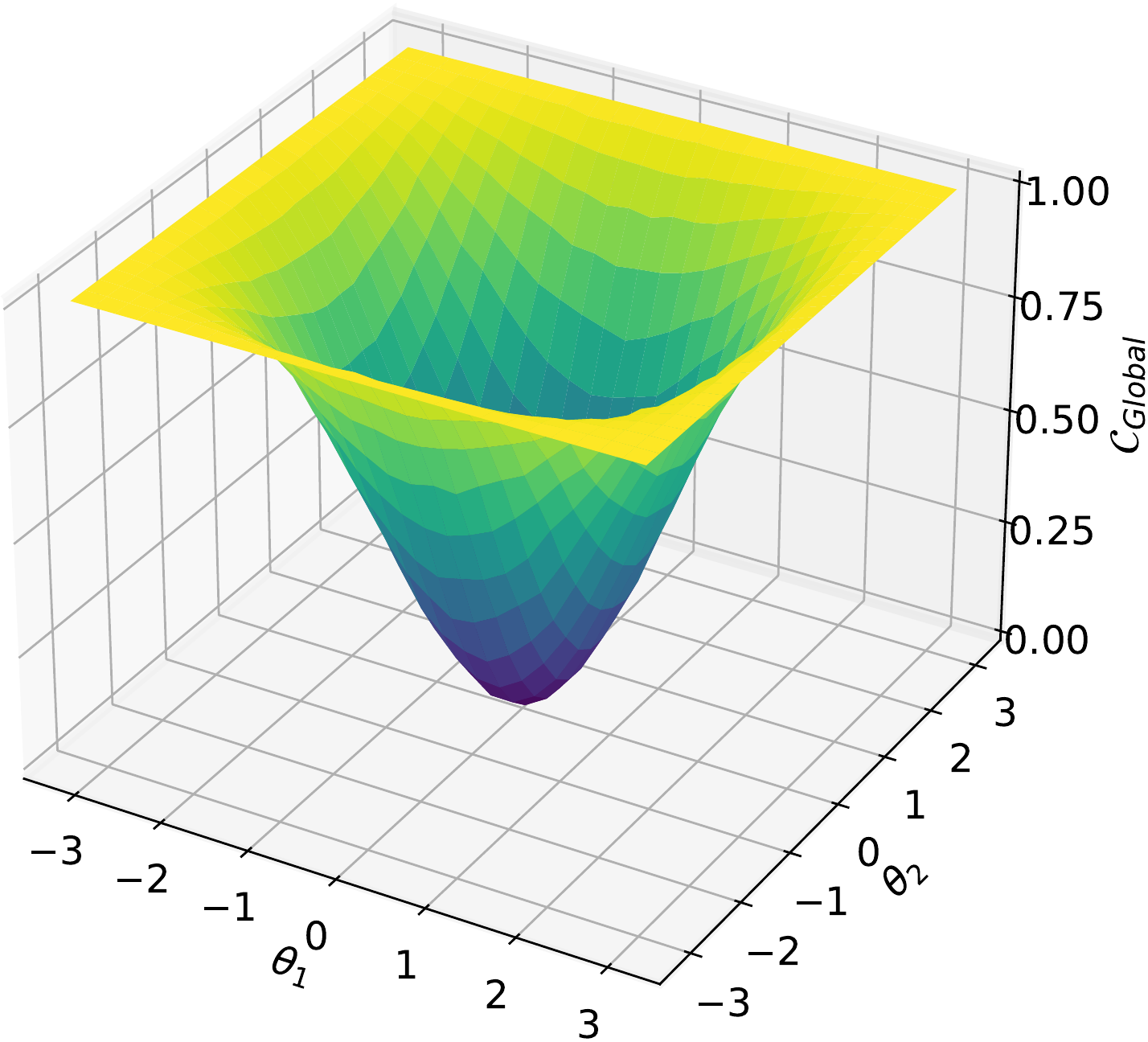}
    \end{minipage}
    \end{subfigure}
    \begin{subfigure}[b]{0.48\linewidth}
    \begin{minipage}{.08\textwidth}
        \caption{}
        \label{fig:barren-plateau-2}
    \end{minipage}%
    \begin{minipage}{0.9\textwidth}
        \includegraphics[width=0.9\textwidth]{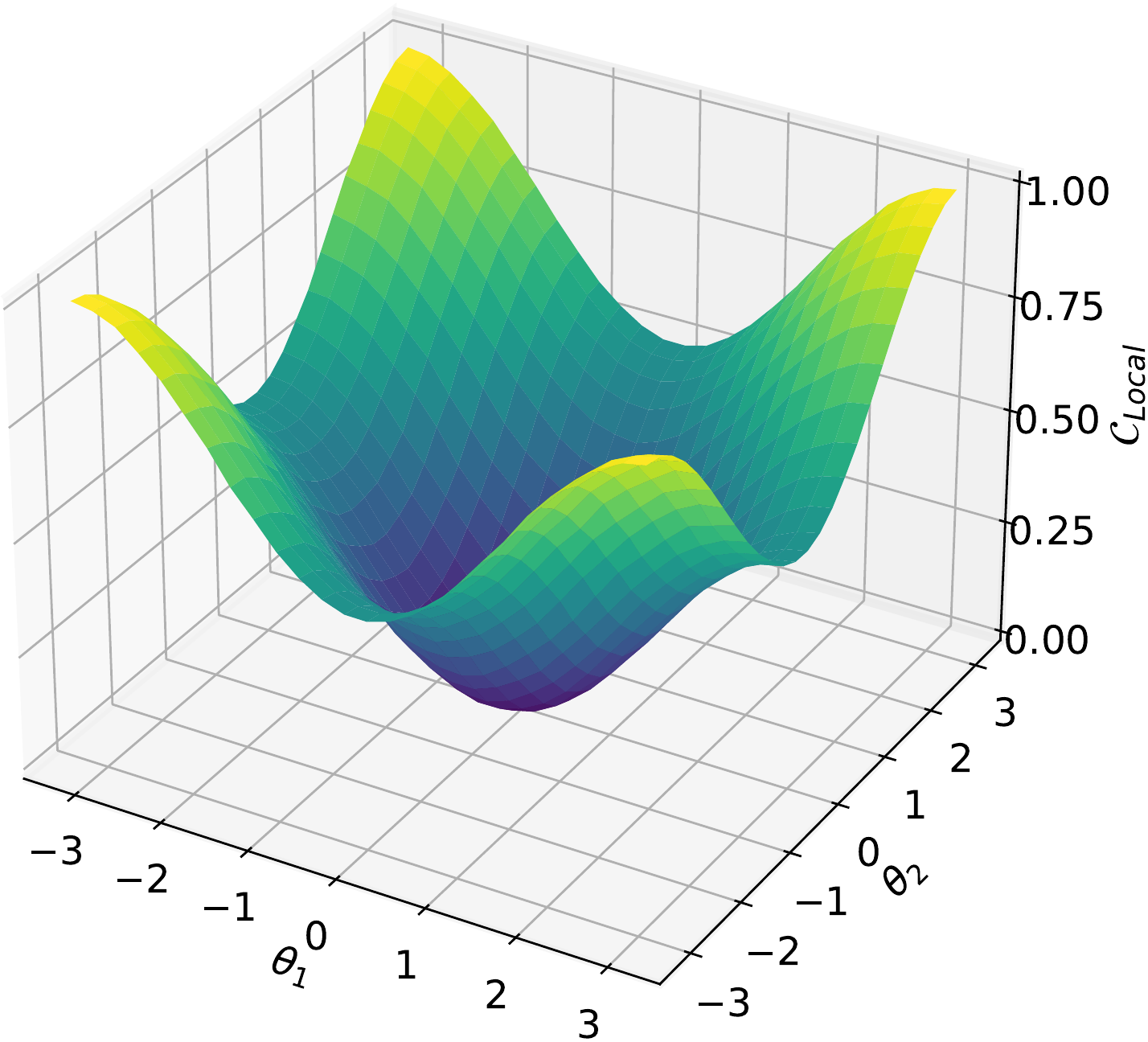}
    \end{minipage}
    \end{subfigure}\\
    \begin{subfigure}[b]{0.48\linewidth}
    \begin{minipage}
    {.03\textwidth}
        \caption{}
        \label{fig:barren-plateau-3}
    \end{minipage}%
    \begin{minipage}{0.90\textwidth}
        \includegraphics[width=.9\linewidth]{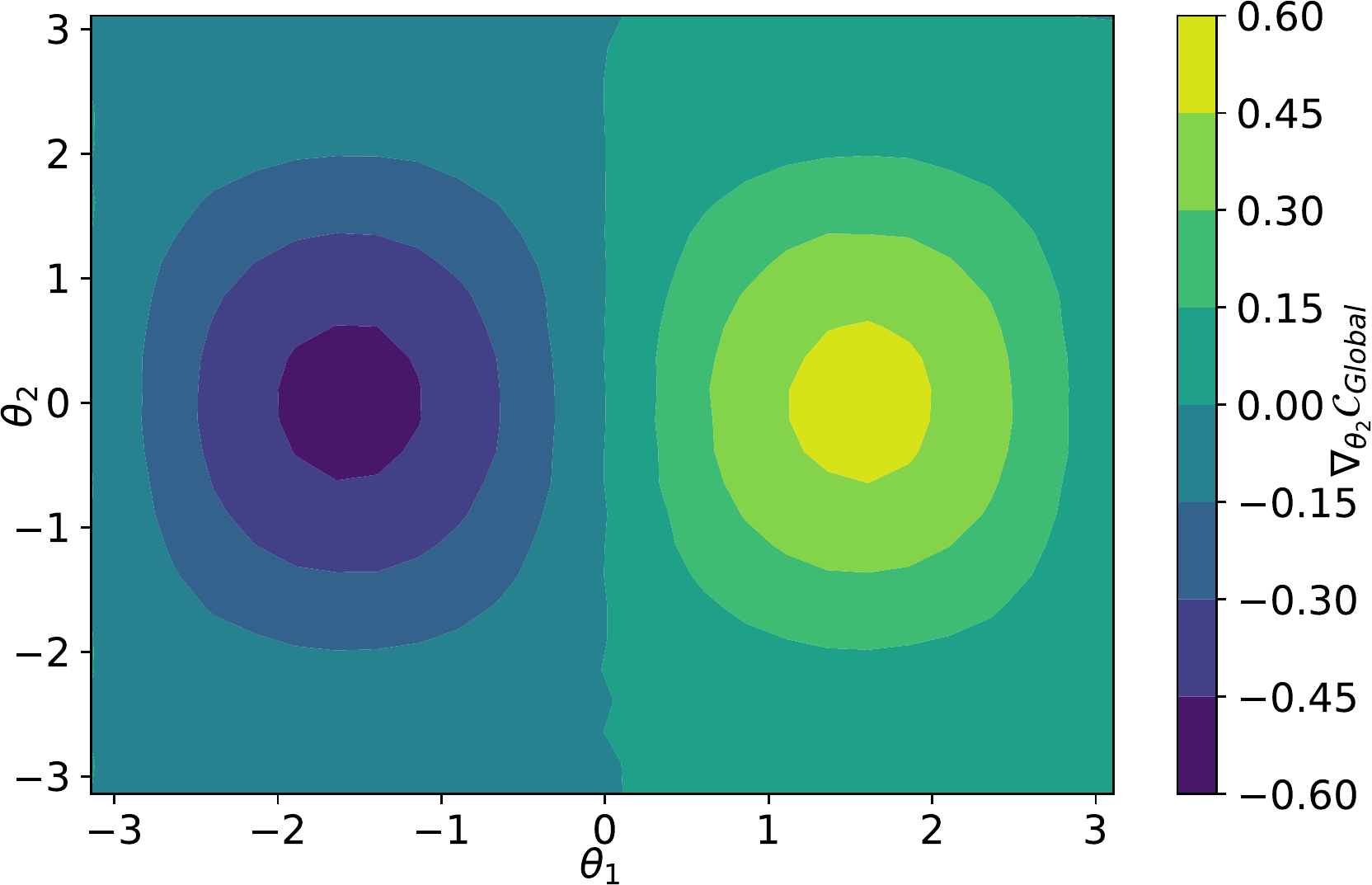}
    \end{minipage}
    \end{subfigure}
    \begin{subfigure}[b]{0.48\textwidth}
    \begin{minipage}{.08\textwidth}
        \caption{}
        \label{fig:barren-plateau-4}
    \end{minipage}%
    \begin{minipage}{0.9\textwidth}
        \includegraphics[width=0.9\textwidth]{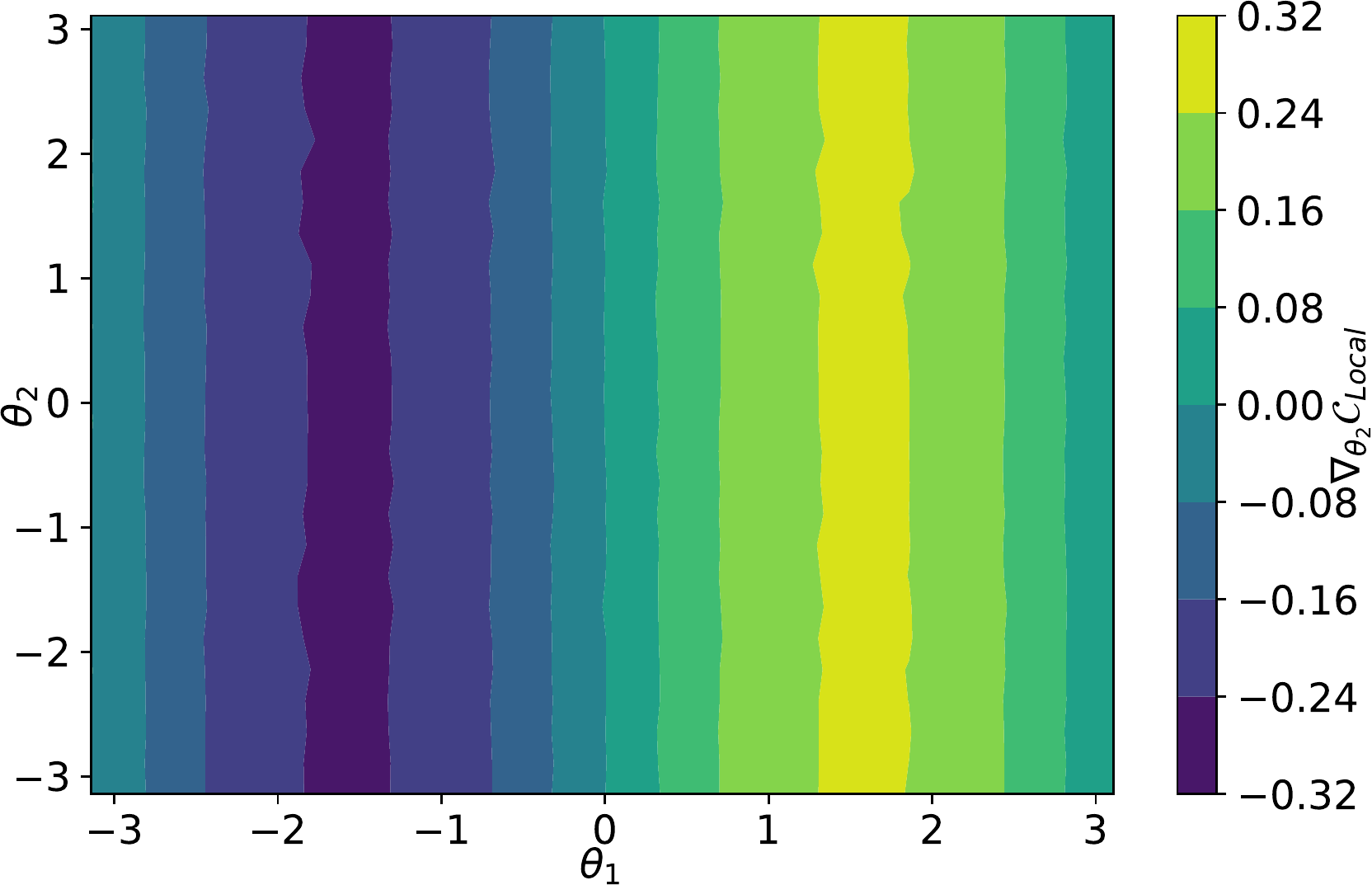}
    \end{minipage}
    \end{subfigure}
    \caption[Presence of barren plateaus in parameterized quantum circuits]{Here we show the emergence of barren plateaus in the task of learning an Identity gate using the ansatz $R_X(0,\theta_1)R_X(1, \theta_2)CZ(0, 1)$ solely based on the choice of the cost function. Figures (a) and (b) represents the loss landscape for the $\mathcal{C}_{Global}$ and local $\mathcal{C}_{Local}$ cost functions, respectively. Similarly, figures (c) and (d) represents coloured heat maps for their  corresponding gradients $\nabla_{\theta_2}\mathcal{C}_{\text{Global}}$ and $\nabla_{\theta_2}\mathcal{C}_{\text{Local}}$} 
    \label{fig:barren-plateau}
\end{figure*}


\subsection{Parameter Histograms}

For our $M$-parameter PQC $\hat{U}(\vec{\theta})$, the parameters $\theta_i$ at the start of the training process are sampled from some prior probability distribution $\pi_0(\theta)$. Through the training process, we desire to learn an optimized join probability distribution over the parameters $\pi^*(\theta)$. This learnt parameter distribution
\begin{equation}
    \pi^* = \underset{\pi(\theta)}{\operatorname{argmin}} \; \underset{\theta \sim \pi}{\mathbb{E}} \mathcal{C}(\vec{\theta}) = \underset{\pi(\theta)}{\operatorname{argmin}} \; \underset{\theta \sim \pi}{\mathbb{E}} Tr[O \hat{U}(\vec{\theta}) \rho \hat{U}^\dagger(\vec{\theta})]
\end{equation}

The evolution of the parameter distribution from $\pi_0 \rightarrow \pi_t \rightarrow \pi^*$ is visualized by our parameter histogram module. The probability distributions are analyzed by starting with an ensemble of vectors $\vec{\theta_l} \sim \pi_0$, letting the entire ensemble evolve using our classical optimization subroutine, and sampling the vectors in the ensemble to get the distribution over parameters at time $t$ as $\pi_t(\vec{\theta})$. 

The marginal distribution over each variable $\pi_t(\vec{\theta_i})$ is plotted at each timestep. Change in the profile of this distribution over consecutive timesteps implies a role of those parameters in those timesteps of the learning process.

\section{\label{sec:challenges}Challenges}

In this section, we will discuss some key challenges that we come across in variational quantum computation and possible ways to identify and mitigate these problems by using tools provided in qLEET.

\subsection{Effect of Noise}

The quantum hardware that exists today are imperfect, as a result of which a computation being run on them may suffer various kinds of errors \cite{Chaudhary2022-kl}. Therefore, in order to realistically simulate and characterize the performance of a parameterized quantum circuit (PQC), we must include these errors in our computation. Our library does so by using noise models from libraries such as Cirq and Qiskit, which provides for errors related to coherent gate errors, incoherent errors, and state preparation and measurement (SPAM) errors. Users can provide the \texttt{NoiseModel} to the \texttt{CircuitSimulator} function in the simulator module while running the experiments. 

Another source of error in quantum computation arises from the limited number of times the circuit is repeatedly executed for sampling. This restricts the precision with which one can compute the Pauli observable $\hat{O}$ for calculating the cost function $\mathcal{C}$ as the number of measurements $m$ required for estimating the expectation value $\langle\hat{O}\rangle$ with precision $\epsilon$ would be $O(1/\epsilon^2)$ \cite{Higgott2019variationalquantum}. In qLEET, the default value of the number of repetitions is $1024$ and is determined by the \texttt{shots} variable, which can be provided at the time of calling any analysis function from the analyzer module.

\vspace{-2pt}

\subsection{Presence of Barren Plateaus}

The main crux of the discussion presented in the previous section is that the choice of ansatz and the cost function together is crucial for successfully training a PQC for a given task. One of the critical hindrances for the training to go as expected is the barren plateau (BP) phenomenon, where the partial derivatives $\partial_{\theta_k}\mathcal{C}(\vec{\theta})$ of the cost function $\mathcal{C}(\vec{\theta})$ with respect to variational parameters $\theta_k$ will, on average, exponentially vanish (Eq. \ref{eq:barren-plateau}). This leads to the flattening of the loss landscape, traversing through, which would require an exponentially large number of shots (for more precision) against finite sampling noise to determine the direction that minimizes the cost. Moreover, it was recently shown in \cite{2020arXiv200714384W} that BPs can also be induced due to noise present in the quantum hardware. This could be a significant issue since it could erase the potential computation advantage associated with quantum computation due to the exponential scaling required to attain the necessary precision, making the complexity comparable to classical algorithms.
\begin{equation}\label{eq:barren-plateau}
	\text{Var}_{\vec{\theta}}[\partial_{\theta_k}\mathcal{C}(\vec{\theta})] \in O\left(\frac{1}{m^N}\right),\quad \text{for}\ m > 1
\end{equation}
In qLEET, one can potentially visualize the BP phenomena by visualizing the loss landscape for a chosen PQC and cost function. This could allow users to see if BP can be mitigated by tweaking either the structure of PQC itself or just the cost function. For example, in Fig. \ref{fig:barren-plateau}, we show an example of BP dependent on the cost function in a shallow ansatz \cite{s41467-021-21728-w}. Here we compare global $\mathcal{C}_{\text{Global}}$ and local $\mathcal{C}_{\text{Local}}$ cost functions for learning the Identity gate using a very simple ansatz: $R_X(0,\theta_1)R_X(1, \theta_2)CZ(0, 1)$. 
\begin{equation}
\begin{split}
    \mathcal{C}_{\text{Global}} &= \bra{\psi(\vec{\theta})} (I - \ket{0\ldots0}\bra{0\ldots0}) \ket{\psi(\vec{\theta})} \\
    &= 1 - p_{0\ldots0}
\end{split}
\end{equation}
\begin{equation}
\begin{split}
    \mathcal{C}_{\text{Local}} &= \bra{\psi(\vec{\theta})} \Bigg(I - \frac{1}{n}\sum_j \ket{0}\bra{0}_j\Bigg) \ket{\psi(\vec{\theta})} \\
    &= 1 - \frac{1}{n}\sum_j p_{0_j}
\end{split}
\end{equation}
We see how the loss landscape flattens for the $\mathcal{C}_{\text{Global}}$ and the gradients vanish exponentially in comparison to  $\mathcal{C}_{\text{Local}}$. In terms of the circuit structure, one way to predict the presence of the BP phenomena for an ansatz is to look at how close its expressivity is to that of a unitary 2-design \cite{Harrow2009} or whether it exhibits excess entanglement that could hinder its trainability \cite{arxiv.2010.15968}. To mitigate BP in such cases require one to restrict the randomness in the circuit by correlating some of the parameters and limit the depth of the circuit by reducing the number of layers, if possible. However, the optimal trade-off between the circuit's trainability and its ability to use quantum resources as quantified by its expressibility and entangling power depends on the nature of the problem and requires a well-designed architecture. For example, for studying spins systems, one might be able to tensor-network based ansatz structure which allows high trainability with sufficient expressibility if its depth is maintained to be shallow \cite{Pesah2021, 2011.06258, Sahoo2022}. In addition to the BP phenomena, we also notice the narrow gorge phenomena, where global minima are contained in a steeply deep valley. This makes it difficult for gradient-based optimization to reach the global minima since it might not have a low learning rate to not overstep inside the gorge. 

\subsection{Estimation of Reachability}

Reachability quantifies whether a given PQC, $\hat{U}(\vec{\theta})$, with parameters $\vec{\theta}$ is capable of representing a parameterized quantum state $\ket{\psi(\vec{\theta})}$ that minimizes the cost function $\mathcal{C}$. Mathematically it is defined as \cite{PhysRevLett.124.090504}:

\begin{equation}
f_\text{R}=\text{min}_{\psi\in\mathcal{H}}\bra{\psi}\mathcal{C}\ket{\psi}-\text{min}_{\vec{\theta}}\bra{\psi(\vec{\theta})}\mathcal{C}\ket{\psi(\vec{\theta})},
\end{equation}

where the first and second term is the minimum over all states $\ket{\psi}$ sampled from the Haar measure and all states that the PQC can represent, respectively. The reachability is equal or greater than zero $f_\text{R}\ge0$, with $f_\text{R}=0$ when the PQC can generate an optimal state $\ket{\psi(\vec{\theta}^*)}$ that minimizes the objective function. This can be easily implemented in qLEET using the \texttt{CircuitSimulator} function present in the simulator module.  

\section{\label{sec:conclusion}Conclusion}

This paper presents an open-source library called qLEET and demonstrates its ability to analyze various properties of parameterized quantum circuits (PQCs), such as their expressibility and entangling power. We motivate the importance of studying these properties from the problem of trainability of PQCs. We have discussed and showed how important insights could be gained from visualizing loss landscapes and training trajectories for variational quantum computation. We also present the theory of expressibility and entangling capability of a PQC based on the deviation of the distribution of parameterized states produced from the Haar measure, which samples uniformly from the entire Hilbert space. We also describe the idea of the entanglement spectrum, which allows visualizing the previous two properties at once. Overall, we demonstrate how different modules included in \texttt{qleet} can be used by users to study various variational algorithms and quantum machine learning models. Finally, we discuss some critical challenges for variational quantum algorithms such as Barren Plateaus and Reachability. We conclude that qLEET will provide opportunities for the quantum community to design new hybrid algorithms by utilizing intuitive insights from the ansatz capability and structure of the loss landscape.

\section*{Data Availability}
The code created to run the presented simulations and any related supplementary data could be made available to any reader upon reasonable request.

\section*{Acknowledgements}
We acknowledge the help and financial support of the Unitary Fund for this project. We also acknowledge Prof. Harjinder Singh for the fruitful discussions we had with him throughout the development of this library.

\section*{Declarations}
The authors have no competing interests to declare relevant to this article's content.

\bibliographystyle{apsrev4-2}

\bibliography{qleet}

\end{document}


\preprint{APS/123-QED}
\title{Supplementary: qLEET - Visualizing Loss Landscapes, Expressibility, Entangling power and Training Trajectories for Parameterized Quantum Circuits}

\author{Utkarsh Azad}
\email{utkarsh.azad@research.iiit.ac.in}
\thanks{Corresponding Author}
\author{Animesh Sinha}
\email{animesh.sinha@research.iiit.ac.in}
\affiliation{
    Center for Computational Natural Sciences and Bioinformatics, International Institute of Information Technology, Hyderabad.\\
    Center for Quantum Science and Technology,\\ International Institute of Information Technology, Hyderabad.
}
\date{\today}

\maketitle

\section{\label{sec:supl-expressibility-tutorial}Tutorial: Entaglement Ability Analysis}

In this section, we will learn how to calcualte expressibility of Parameterized Quantum Circuits (PQCs) using qLEET, which could thought of as traversing power of a PQC in the Hilbert space. We look at different parameterized states generated by the sampled ensemble of parameters for a given PQC. We then compare the resulting distribution of state fidelities ($\mathcal{F}$) generated by this sampled ensemble to that of the ensemble of Haar random states.

We currently support two expressibility measures - \textbf{Kullback–Leibler Divergence} and \textbf{Jensen–Shannon Divergence}
$$ \textrm{Expressibility} = D_{\textrm{KL}} \Big( \hat{P}_{\textrm{PQC}}(\mathcal{F}; \theta) \big\vert P_{\textrm{Haar}}(\mathcal{F}) \Big) $$
$$ \textrm{Expressibility} = D_{\sqrt{\textrm{JSD}}} \Big( \hat{P}_{\textrm{PQC}}(\mathcal{F}; \theta) \big\vert P_{\textrm{Haar}}(\mathcal{F}) \Big) $$

All circuit analysis using qleet begins with defining a parameterized quantum circuit using a library of choice, and then passing it into qleet's \lstinline{CircuitDescriptor} interface.

\begin{lstlisting}
params = [qiskit.circuit.Parameter(r"$\theta_1$")]

qiskit_circuit = qiskit.QuantumCircuit(1)
qiskit_circuit.h(0)
qiskit_circuit.rz(params[0], 0)
qiskit_descriptor = qleet.interface.circuit.CircuitDescriptor(
    circuit=qiskit_circuit, params=params, cost_function=None
)
\end{lstlisting}

The analyze the expressibility, we can use the corresponding analyzer. We can get the expressibility using either of the two supported measures.

\begin{lstlisting}

qiskit_expressibility = qleet.analyzers.expressibility.Expressibility(
    qiskit_descriptor, samples=100
)
expr_jsd = qiskit_expressibility.expressibility("jsd")
print("JSD Expressibility:", expr_jsd)

expr_kld = qiskit_expressibility.expressibility("kld")
print("KLD Expressibility:", expr_kld)

plt_figure = qiskit_expressibility.plot()
\end{lstlisting}

We look at different parameterized states generated by the sampled ensemble of parameters for a given PQC. We then compare the resulting distribution of eigenvalues of the bipartite state generated by this sampled ensemble to that of the ensemble of eigenvalues of Haar random states.

We currently support two measures to calculate entanglement spectrum divergence (ESD) - \textbf{Kullback–Leibler Divergence} and \textbf{Jensen–Shannon Divergence}
$$\textrm{ESD} = D_{\textrm{KL}} \Big(\hat{P}_{\textrm{PQC}}(H_{\textrm{ent}}; \theta) \big\vert P_{\textrm{Haar}}(H_{\textrm{ent}}) \Big) $$
$$\textrm{ESD} = D_{\sqrt{\textrm{JSD}}} \Big(\hat{P}_{\textrm{PQC}}(H_{\textrm{ent}}; \theta) \big\vert P_{\textrm{Haar}}(H_{\textrm{ent}}) \Big) $$

\begin{lstlisting}
params = [
    qiskit.circuit.Parameter(r"$\theta_1$"),
    qiskit.circuit.Parameter(r"$\theta_2$")
]
qiskit_circuit = qiskit.QuantumCircuit(2)
qiskit_circuit.rx(params[0], 0)
qiskit_circuit.cx(0, 1)
qiskit_circuit.rx(params[1], 1)
qiskit_descriptor = qleet.interface.circuit.CircuitDescriptor(
    circuit=qiskit_circuit, params=params, cost_function=None
)
\end{lstlisting}

\begin{lstlisting}
analyzer = (
    qleet.analyzers.entanglement.EntanglementCapability(
        qiskit_descriptor, samples=500
    )
)

entanglement_mw = analyzer.entanglement_capability("meyer-wallach")
print("Entanglement Capability (Meyer Wallach Measure):", entanglement_mw)

entanglement_scott = analyzer.entanglement_capability("scott")
print("Entanglement Capability (Scott Measure):", entanglement_scott)
\end{lstlisting}

In this section, we will plot the entanglement spectrum.

\begin{lstlisting}
def ansatz(params, cparams=None):
    layers, num_qubits, depth = params.shape
    ansatz = qiskit.QuantumCircuit(num_qubits)
    for idx in range(layers):
        if idx:
            ansatz.barrier()
        for ind in range(num_qubits):
            ansatz.rx(params[idx][ind][0], ind)
            ansatz.rz(params[idx][ind][1], ind)
            ansatz.rx(params[idx][ind][2], ind)
        for ind in range(num_qubits-1):
            ansatz.cx(ind, ind+1)
    return ansatz
\end{lstlisting}

\begin{lstlisting}
data = []
results = []
num_qubits = 12
for idx in range(1, 17):
    print(idx, end=' ')
    params = np.array([qiskit.circuit.Parameter(fr"$\theta_{idx}$")
                       for idx in range(idx*num_qubits*3)])
    qiskit_descriptor = qleet.CircuitDescriptor(
        circuit=ansatz(np.array(params).reshape((idx, num_qubits, 3))), 
        params=params, cost_function=None
    )
    qiskit_entanglement_spectrum = \
        qleet.analyzers.entanglement_spectrum.EntanglementSpectrum(
            qiskit_descriptor, samples=100
        )
    pqc_esd, mean_eig = qiskit_entanglement_spectrum.entanglement_spectrum("jsd")
    results.append(pqc_esd)
    data.append(mean_eig)
data = np.array(data)

fig = qiskit_entanglement_spectrum.plot(data)
\end{lstlisting}

\section{\label{sec:supl-loss-tutorial}Loss Landscape and Training Trajectory Analysis}

For this section of the tutorial, we shall be constructing our circuits in the \textbf{Cirq} library, which is also supported by our multi-backend analyzer. Using cirq, we define a parameterized quantum circuit, we define its parameters as sympy symbols, and we define a cost function as a Pauli measurement on the outputs of this circuits. All of this is passed into out \lstinline{CircuitDescriptor} interface

\begin{lstlisting}
graph = nx.gnm_random_graph(n=8, m=20)
qubits = cirq.GridQubit.rect(1, graph.number_of_nodes())
p = 5

params = sympy.symbols("q0:%d" % (2 * p))
qaoa_circuit = cirq.Circuit()
for qubit in qubits:
    qaoa_circuit.append(cirq.H(qubit))
for i in range(p):
    for edge in graph.edges():
        qaoa_circuit += cirq.CNOT(qubits[edge[0]], qubits[edge[1]])
        qaoa_circuit += cirq.rz(params[2 * i]).on(qubits[edge[1]])
        qaoa_circuit += cirq.CNOT(qubits[edge[0]], qubits[edge[1]])
    for j in range(len(qubits)):
        qaoa_circuit += cirq.rx(2 * params[2 * i + 1]).on(qubits[j])

qaoa_cost = cirq.PauliSum()
for edge in graph.edges():
    qaoa_cost += cirq.PauliString(1 / 2 * cirq.Z(qubits[edge[0]]) * 
                                  cirq.Z(qubits[edge[1]]))

circuit = qleet.interface.circuit.CircuitDescriptor(
    qaoa_circuit, params, qaoa_cost)
solver = qleet.simulators.pqc_trainer.PQCSimulatedTrainer(circuit)
\end{lstlisting}

\begin{lstlisting}
class MaxCutMetric(qleet.interface.metric_spec.MetricSpecifier):

    def __init__(self, graph):
        super().__init__("samples")
        self.graph = graph

    def from_samples_vector(self, samples_vector):
        return np.mean([nx.algorithms.cuts.cut_size(
            self.graph, np.where(cut)[0]) for cut in samples_vector])
    
    def from_density_matrix(self, density_matrix):
        raise NotImplementedError
    
    def from_state_vector(self, state_vector):
        raise NotImplementedError

metric = MaxCutMetric(graph)
\end{lstlisting}

\begin{lstlisting}
plot = qleet.analyzers.loss_landscape.LossLandscapePlotter(
    solver, metric, dim=2)
solver.train(n_samples=5000)
fig_loss_surface = plot.plot("surface", points=20)

trackers = qleet.interface.metas.AnalyzerList(
    qleet.analyzers.training_path.LossLandscapePathPlotter(plot),
    qleet.analyzers.training_path.OptimizationPathPlotter(mode="tSNE"),
)
for _i in range(5):
    solver.train(loggers=trackers, n_samples=5000)
    trackers.next()
    
fig_loss_traversal = trackers[0].plot()
fig_training_trace = trackers[1].plot()
\end{lstlisting}

\begin{figure*}[htp]
    \centering
    \begin{subfigure}[b]{0.32\linewidth}
        \includegraphics[width=\textwidth]{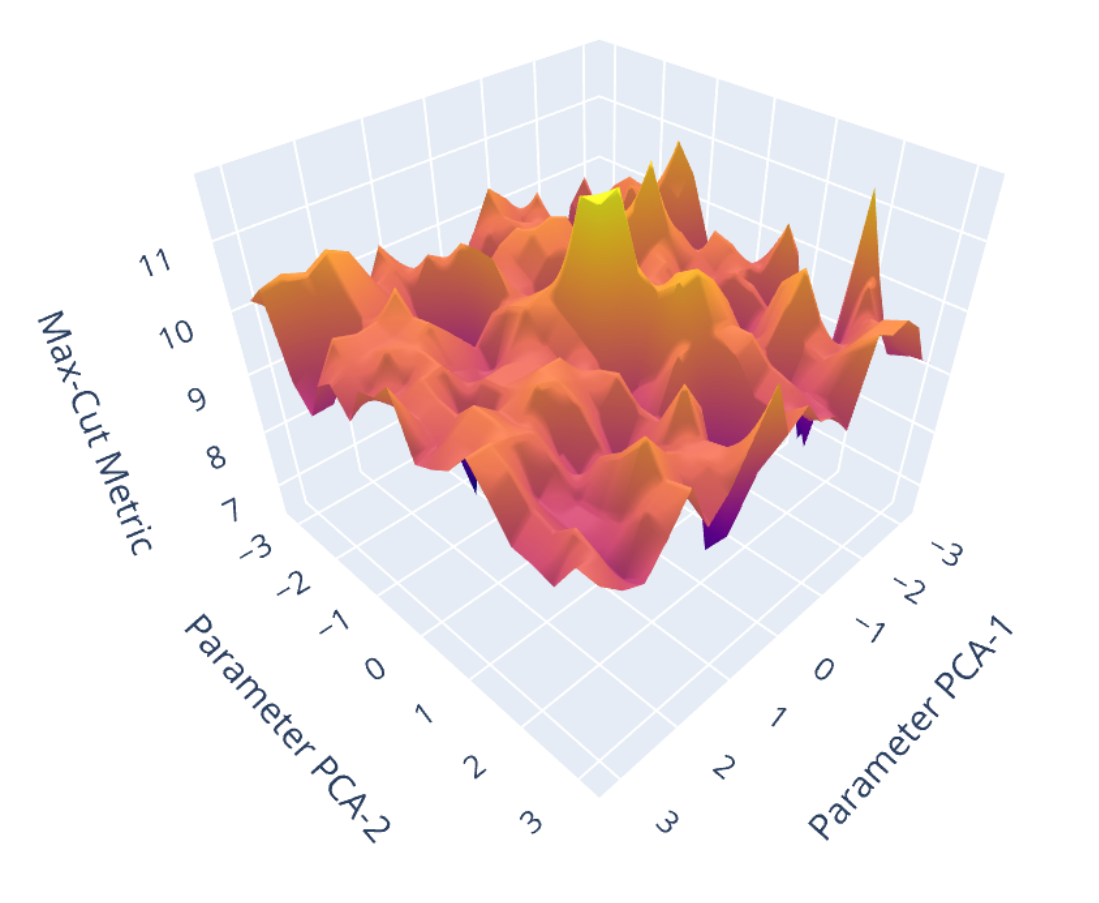}
        \caption{Metric Landscape (inverse of loss) around obtained optima}
    \end{subfigure}
    \begin{subfigure}[b]{0.32\linewidth}
        \includegraphics[width=\textwidth]{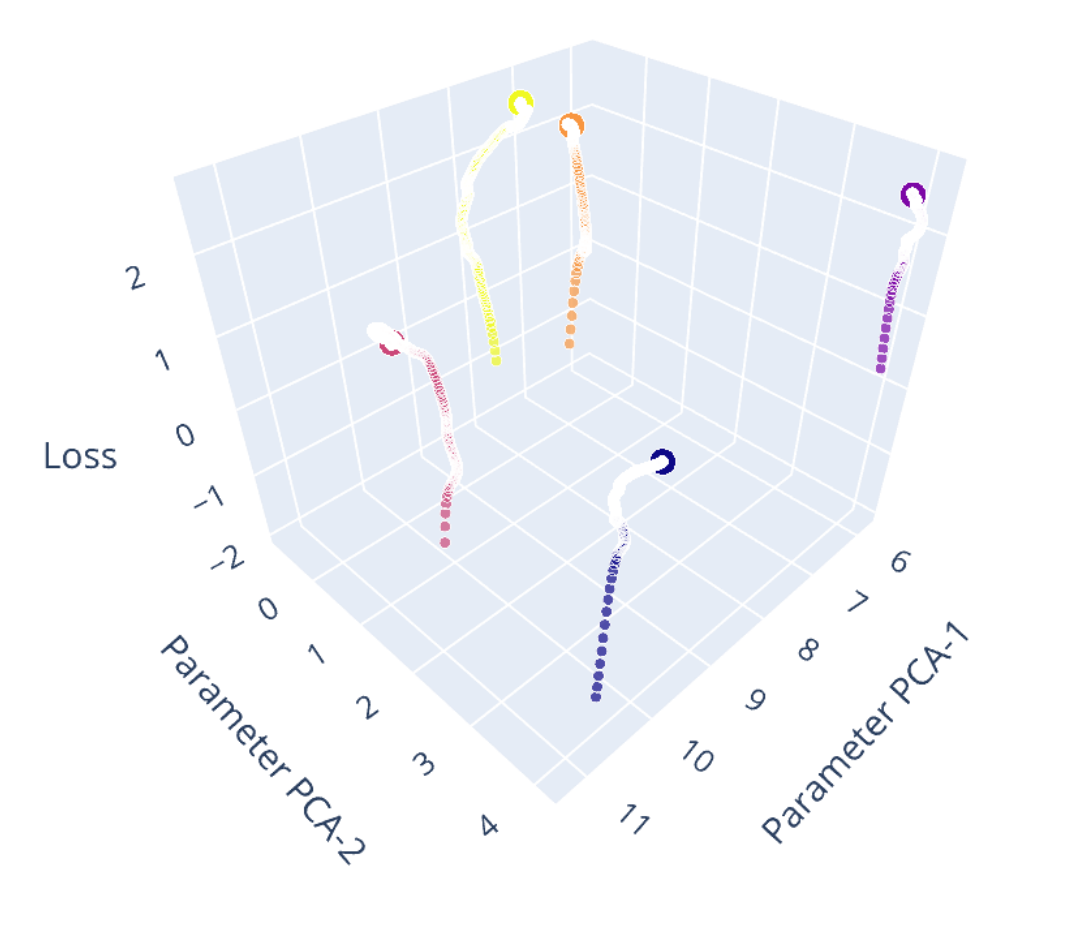}
        \caption{PCA plot with loss for training trajectories of 5 runs}
    \end{subfigure}
    \begin{subfigure}[b]{0.32\linewidth}
        \includegraphics[width=\textwidth]{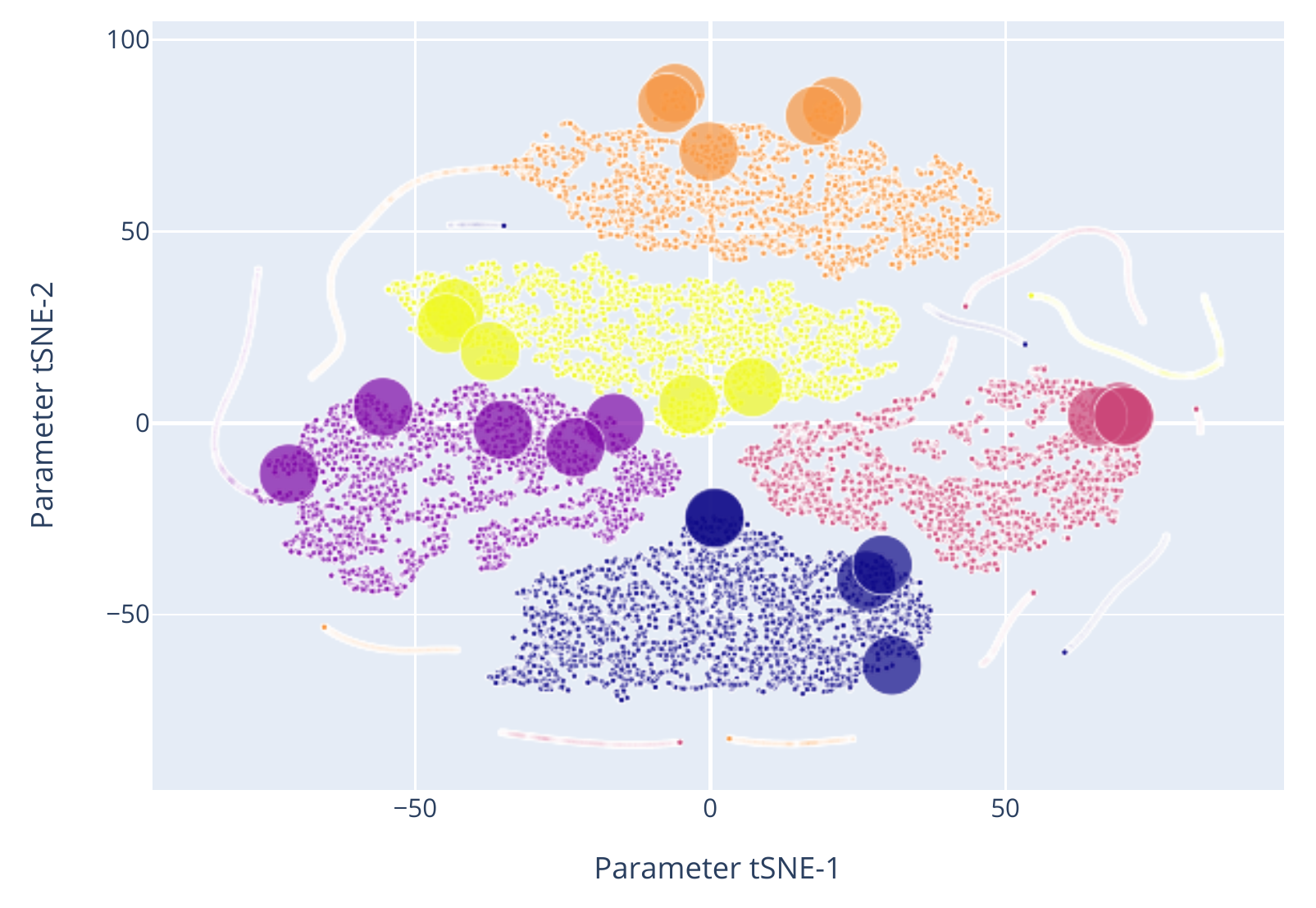}
        \caption{2-D tSNE of training trajectories from 5 runs}
    \end{subfigure}%
    \caption{Loss and Training Trajectory plots obtained on analyzing the circuit shown. Here, the analysis is shown for a circuit representing max-cut on a graph with 8 nodes and 20 edges.}
    \label{fig:loss-land-train-traj}
\end{figure*}

\section{\label{sec:supl-mz-op}Entanglement Analysis for $M_Z$ operator \cite{Zidan2020}}

\begin{figure}
    \centering
    \includegraphics{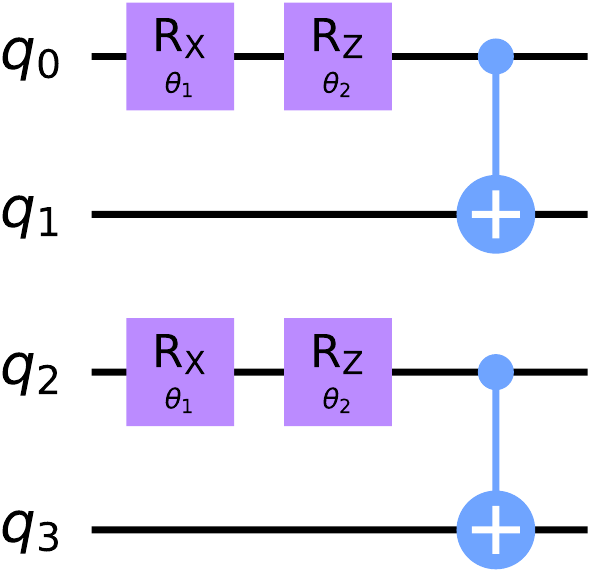}
    \caption{$M_Z$ operator used in the quantum computing model based on entanglement degree allows to differentiate between the non-orthogonal states of the form $e_1|0\rangle + e_2|1\rangle$, with arbitrary accuracy \cite{Zidan2020, Khan2022, Punla2021, Panda2022}.}
    \label{fig:my_label}
\end{figure}

\begin{lstlisting}
params = [qiskit.circuit.Parameter(r"$\theta_1$"), 
          qiskit.circuit.Parameter(r"$\theta_2$"), 
          qiskit.circuit.Parameter(r"$\theta_3$"), 
          qiskit.circuit.Parameter(r"$\theta_4$")]

qiskit_circuit = qiskit.QuantumCircuit(4)
qiskit_circuit.rx(params[0], 0)
qiskit_circuit.rz(params[1], 0)
qiskit_circuit.rx(params[2], 2)
qiskit_circuit.rz(params[3], 2)
qiskit_circuit.cx(0, 1)
qiskit_circuit.cx(2, 3)

qiskit_descriptor = qleet.interface.circuit.CircuitDescriptor(
    circuit=qiskit_circuit, params=params, cost_function=None
)

qiskit_entg_capability = (
    qleet.analyzers.entanglement.EntanglementCapability(
        qiskit_descriptor, samples=1000
    )
)

entanglement_mw = qiskit_entg_capability.entanglement_capability("meyer-Wallach")
# >>> entanglement_mw = 0.5010648894421558

entanglement_scott = qiskit_entg_capability.entanglement_capability("scott")
# >>> entanglement_scott = array([0.4979689 , 0.38654991])


\end{lstlisting}

\section{Quantum Circuits from the Experiments}

\begin{figure}[!ht]
    \centering
    \includegraphics[width=\textwidth]{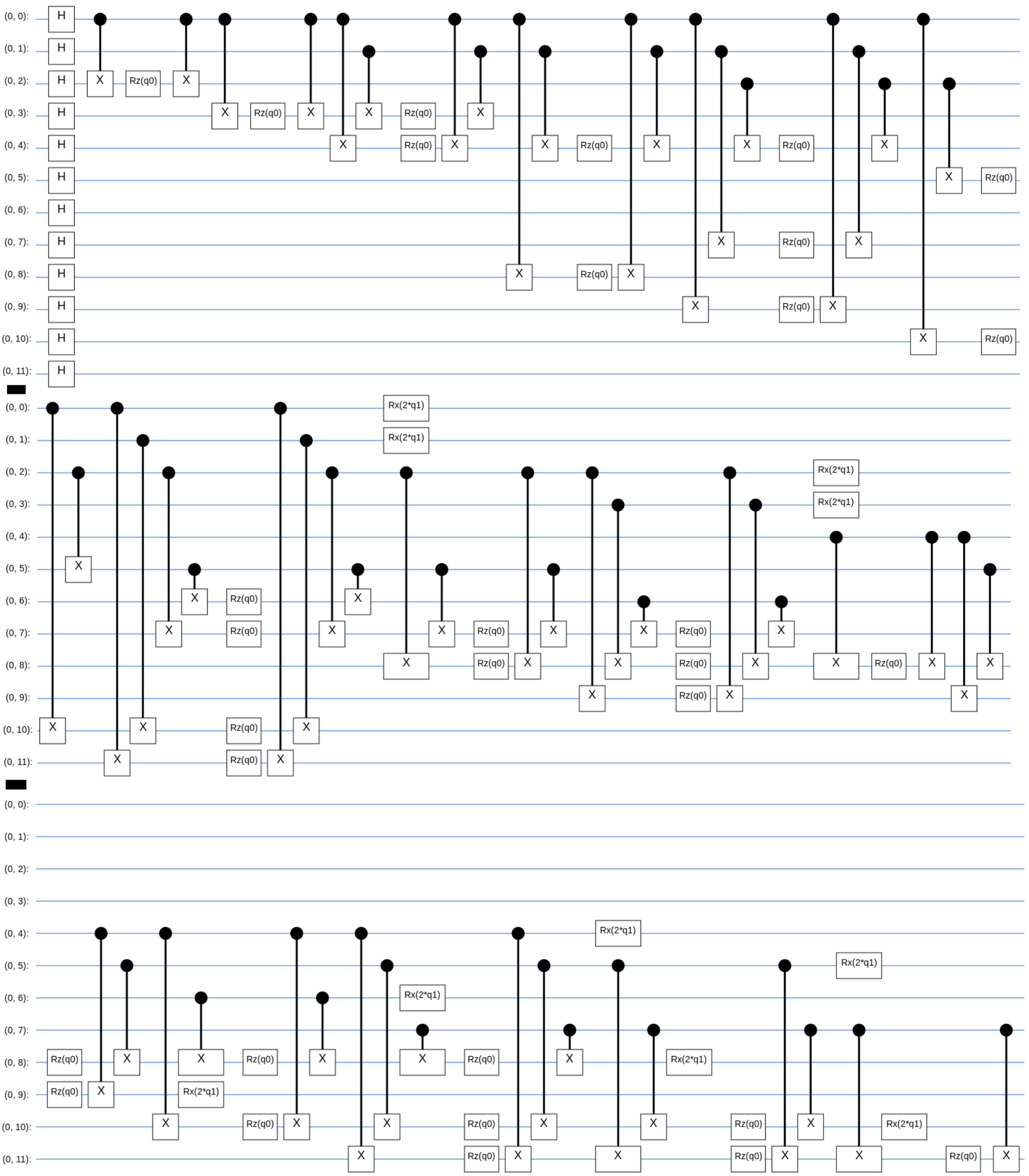}
    \caption{QAOA circuit for p=1. This circuit (except the first Hadamard layer) will be repeated $k$ times for $p=k$.}
    \label{fig:qaoa-circuit}
\end{figure}

\subsection{Loss Landscape and Training Trajectories (Fig. \ref{fig:qaoa-circuit} $\rightarrow$ Fig. 3)}

\begin{figure}[!ht]
    \centering
    \includegraphics[width=0.5\textwidth]{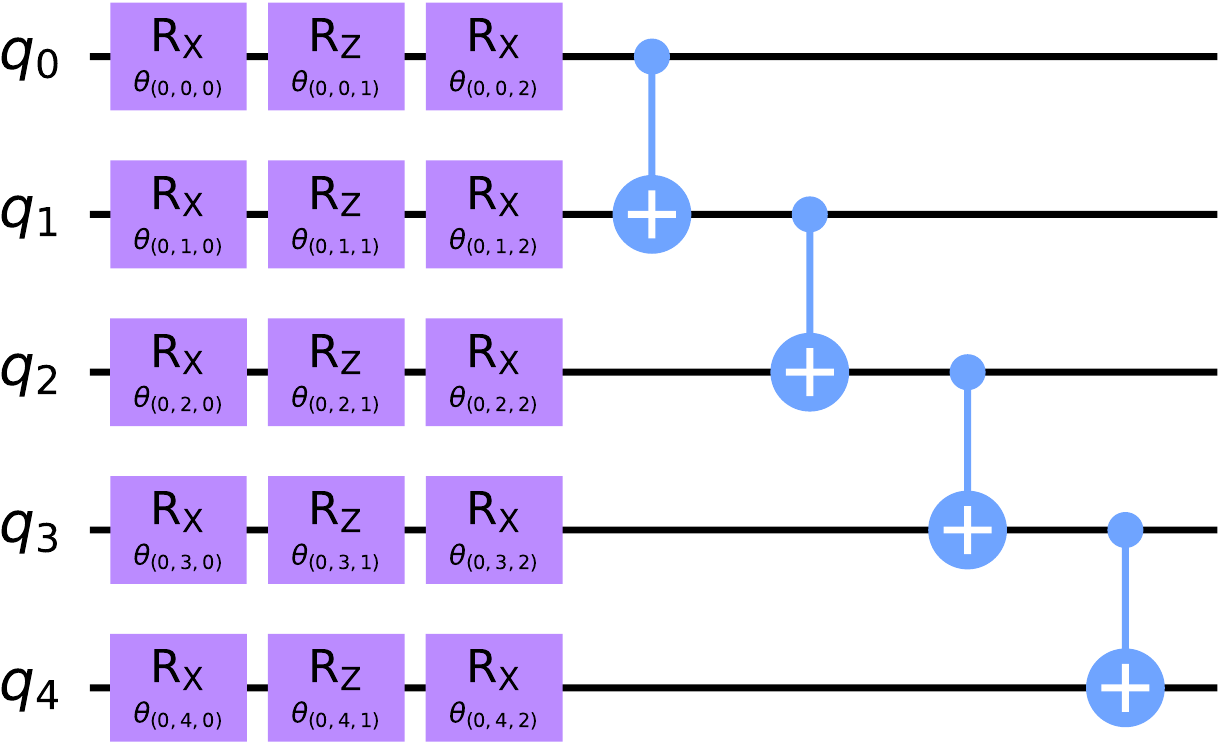}
    \caption{Parameterized quantum circuit $U(\vec{\theta}) =  \prod_{1}^{L}\big(\bigotimes_{i=1}^{5}R_x(\theta_i^1)R_z(\theta_i^2)R_x(\theta_i^3) \ldots \bigotimes_{i<j}CX(i, j)\big)$}
    \label{fig:expr-circuit}
\end{figure}

\subsection{Expressibility (Fig. \ref{fig:expr-circuit} $\rightarrow$ Fig. 5)}

\begin{figure}[!h]
    \centering
    \includegraphics[width=0.25\textwidth]{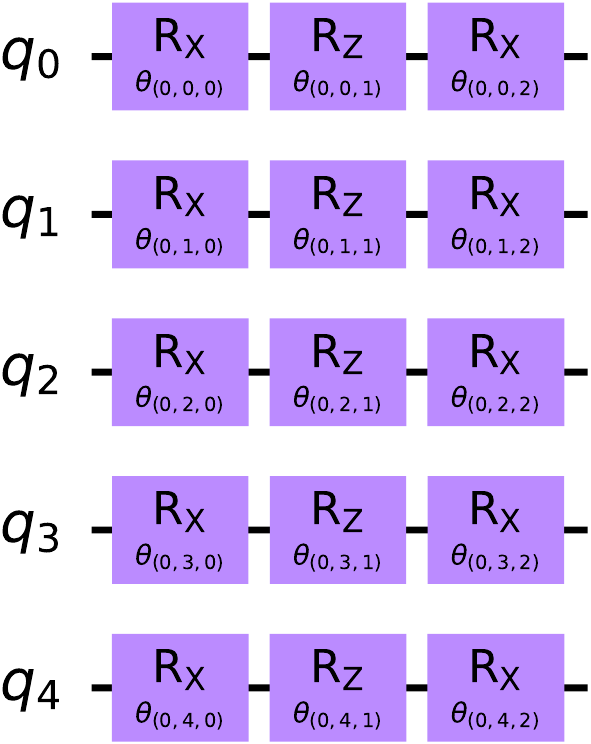}
    \caption{Parameterized quantum circuit $U(\vec{\theta}) = \bigotimes_{i=1}^{5}R_x(\theta_i^1)R_z(\theta_i^2)R_x(\theta_i^3)$}
    \label{fig:entg-circuit}
\end{figure}

\subsection{Entangling Capability (Fig. \ref{fig:entg-circuit} $\rightarrow$ Fig. 6)}

\begin{figure}[!h]
    \centering
    \includegraphics[width=\textwidth]{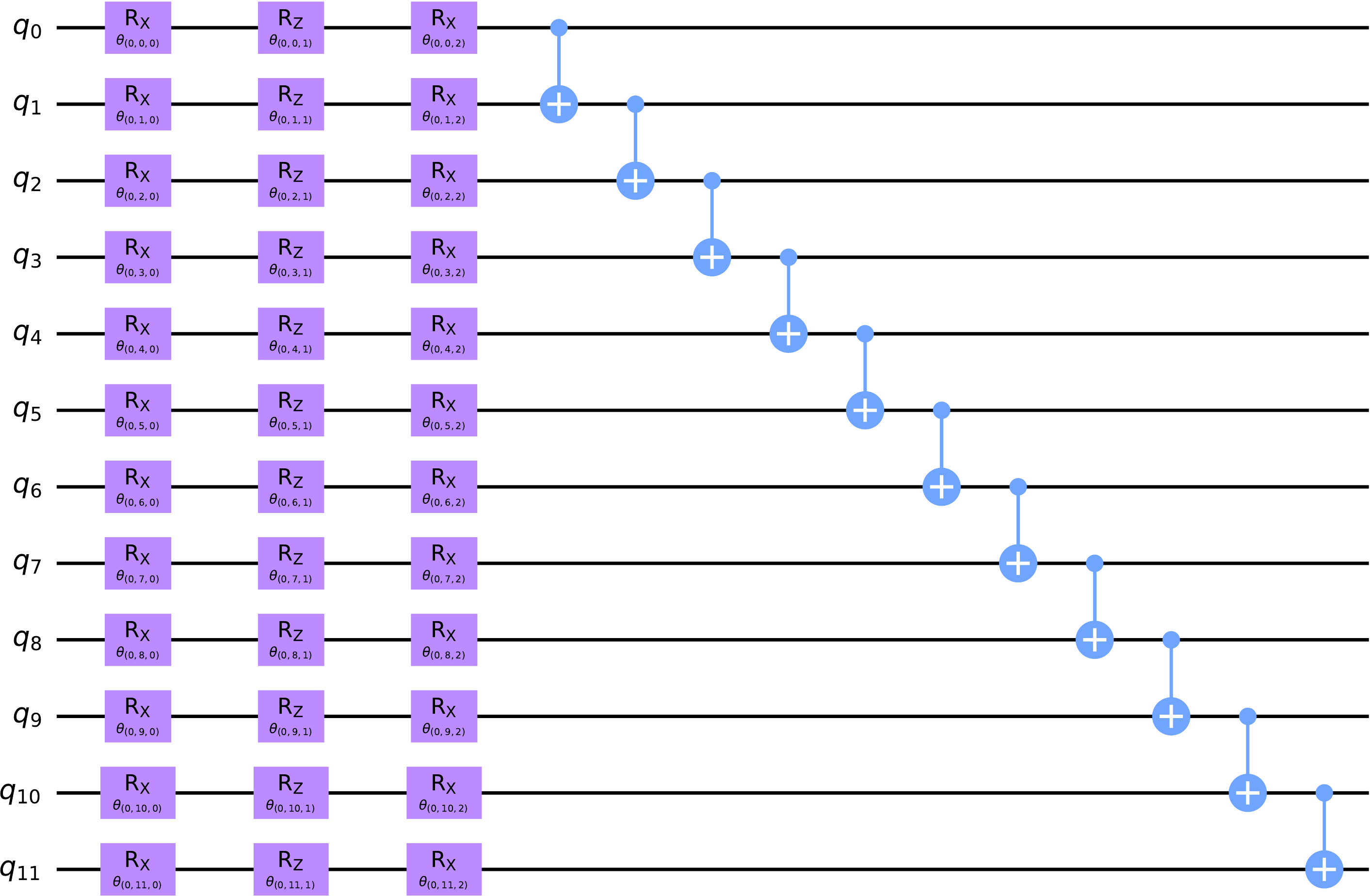}
    \caption{Parameterized quantum circuit $U(\vec{\theta}) = \prod_{1}^{L}\big(\bigotimes_{i=1}^{12}R_x(\theta_i^1)R_z(\theta_i^2)R_x(\theta_i^3) \ldots \bigotimes_{i=1}^{11}CX(i, i+1)\big)$}
    \label{fig:spec-circuit}
\end{figure}

\subsection{Entanglement Spectrum (Fig. \ref{fig:spec-circuit} $\rightarrow$ Fig. 7)}

\bibliographystyle{apsrev4-2}

\bibliography{qleet}